\documentclass[11pt]{article}%{amsart}
\usepackage{amssymb}
\usepackage{citesort}
\usepackage{url}
\usepackage{a4}
\usepackage{cite}
\usepackage{ntheorem}
\newcommand{\sgstatic}{{strictly globally static{}}}
\newcommand{\gstatic}{{globally static{}}}

\newcommand{\riemg}{g}
\newcommand{\riemgz}{g_0}

\newcommand{\<}{\langle}
\renewcommand{\>}{\rangle}

\renewcommand{\hbar}{{\overline \riemgz}}

\newcommand{\bS}{{\overline \Sigma}}
\newcommand{\pS}{{\partial \Sigma}}

\newcommand{\zSigma}{{\mathring{\Sigma}}}

\newcommand{\mLX}{{\mcL_X}}
\newcommand{\mcE}{{\mycal E}}

\newcommand{\lormet  }{{\frak g}}

\newcommand{\nablash}{\nabla{\kern -.75 em
     \raise 1.5 true pt\hbox{{\bf/}}}\kern +.1 em}
\newcommand{\Deltash}{\Delta{\kern -.69 em
     \raise .2 true pt\hbox{{\bf/}}}\kern +.1 em}
\newcommand{\Rslash}{R{\kern -.60 em
     \raise 1.5 true pt\hbox{{\bf/}}}\kern +.1 em}

\newcommand{\mcU}{{\mycal U}}

\newcommand{\hyp}{\Sigma}%{\mycal S}}
\newcommand{\bg}{{\overline{g}_\Sigma}}

\newcommand{\threeg}{{g_\Sigma}} %three dimensional metric pulled back
\newcommand{\mcM}{{\mycal M}}

\newcommand{\bea}{\begin{eqnarray}}
\newcommand{\beaa}{\begin{eqnarray*}}
\newcommand{\bean}{\begin{eqnarray}\nonumber}

\newcommand{\bel}[1]{\begin{equation}\label{#1}}
\newcommand{\beal}[1]{\begin{eqnarray}\label{#1}}
\newcommand{\beadl}[1]{\begin{deqarr}\label{#1}}
\newcommand{\eeadl}[1]{\arrlabel{#1}\end{deqarr}}
\newcommand{\eeal}[1]{\label{#1}\end{eqnarray}}
\newcommand{\eead}[1]{\end{deqarr}}
\newcommand{\eea}{\end{eqnarray}}
\newcommand{\eeaa}{\end{eqnarray*}}

\newcommand{\be}{\begin{equation}}
\newcommand{\ee}{\end{equation}}

\newcommand{\eq}[1]{\eqref{#1}}

\DeclareFontFamily{OT1}{rsfs}{}
\DeclareFontShape{OT1}{rsfs}{m}{n}{ <-7> rsfs5 <7-10> rsfs7 <10->
rsfs10}{} \DeclareMathAlphabet{\mycal}{OT1}{rsfs}{m}{n}
\def\scri{{\mycal I}}%
\def\Scri{\scri}

\usepackage{amsmath} % pour utiliser les caracteres R, Z, C

\let\ssection=\section

\renewcommand{\section}{\setcounter{equation}{0}\ssection}

\newtheorem{defi}{\sc Definition\rm}[section]

\newtheorem{theorem}[defi]{\sc Theorem\rm}
\newtheorem{Theorem}[defi]{\sc Theorem\rm}

\newtheorem{proposition}[defi]{\sc Proposition\rm}

\theorembodyfont{\upshape}
\newtheorem{Remark}[defi]{{\sc Remark}\rm}

\newtheorem{remark}[defi]{{\sc Remark}\rm}

\newcommand{\qed}{\hfill $\Box$\bigskip}

\newcommand{\proof}{\noindent {\sc Proof:\ }}

\def \Reel{\mathbb{R}}

\def \R {\Reel}

\newcommand{\mcL}{{\mycal L}}

\def \Z{\mathbb{Z}}

\newcounter{mnotecount}[section]

\newcommand{\rmnote}[1]{}%{\mnote{#1}}

\begin{document}
\title{Non-trivial,  static, geodesically complete, vacuum space-times with
a negative cosmological constant}
\author{
Michael T. Anderson\thanks{Partially supported by NSF Grant DMS
0072591; email \protect\url{anderson@math.sunysb.edu}}
\\
%\address{
Department of Mathematics\\
S.U.N.Y. at Stony Brook\\
Stony Brook, N.Y. 11794-3651\\
\\
%\author%[P. Chru\'sciel]
Piotr T. Chru\'sciel\thanks{Partially supported by a Polish
Research Committee grant; email \protect\url{
piotr@gargan.math.univ-tours.fr}}
\\
Albert Einstein Institute\thanks{Visiting Scientist. Permanent
address: D\'epartement de math\'ematiques, Facult\'e des
Sciences, Parc de Grandmont, F37200 Tours, France}\\
D-14476 Golm, Germany
\\
\\
 Erwann Delay\thanks{Partially supported by the ACI program of the French Ministry of Research; email
\protect\url{delay@gargan.math.univ-tours.fr}}
%\author[E. Delay (DRAFT)]{Erwann Delay}
%\address{
\\ D\'epartement de math\'ematiques \\
Facult\'e des Sciences \\ Parc de Grandmont \\ F37200 Tours,
France}
\date{}

\maketitle

%%%%%%%%%%%%%%%%%%%%%%%%% resume %%%%%%%%%%%%%%%%%%%%%%%%%%%%%%ù

\begin{abstract}
We construct a large class of new singularity-free static
Lorentzian four-dimensional solutions of the vacuum Einstein
equations with a negative cosmological constant. The new families
of metrics contain space-times with, or without, black hole
regions. Two uniqueness results are also established.
\end{abstract}

%\tableofcontents

%\vspace{2cm}
%\noindent

%{\bf Keywords} :
%...
%\\
%\newline

%{\bf 2000 MSC} :
%...

%%%%%%%%%%%%%%%%%%%%%%%%%%%% intro %%%%%%%%%%%%%%%%%%%%%%%%%%%%%%%%

\section{Introduction}\label{Sintro}

 It is part of the folklore expectations in
general relativity that the following statements hold for
solutions of Einstein's equations, with or without a cosmological
constant:
\begin{itemize}
\item Static non-singular solutions possess at least three
linearly independent local Killing vector fields near each point.
\item Stationary non-singular solutions possess at least two
linearly independent local Killing vector fields near each point.
\end{itemize} By local Killing vector fields we mean  those
solutions of the Killing equations which are defined in a
neighborhood of some point, and which do not necessarily extend to
global solutions.\footnote{An example is given by rotational
Killing vector fields on a torus, which exist in a neighborhood of
each point, but which do not extend to globally defined vector
fields. The restriction to local Killing vector fields is
necessary in the statements above: the four dimensional higher
genus Kottler black holes  have four locally defined Killing
vector fields in a some neighborhood of each point, but only one
which is globally defined.}
 There is a wide body of evidence that these
statements are correct when the cosmological constant
vanishes\footnote{In the $\Lambda=0$ case the only exception known
to us is provided by the Myers --- Nicolai-Korotkin metrics
\cite{Myers:1987rx,Korotkin:1994dw} which are, however, not
asymptotically flat in the sense which one usually uses in the
context of black hole space-times.} or is positive
(see~\cite{manderson:static,manderson:stationary,Chstatic,ChAscona,Weinstein3,LafontaineRozoy2}
and references therein), and some very partial results indicating
that this could perhaps be true when the cosmological constant
$\Lambda$ is
negative~\cite{ChHerzlich,ChruscielSimon,Galloway:2001uv,Galloway:2002ai,BGH}.
The object of this paper is to show that such rigidity is false in
this last situation.  More precisely, for $\Lambda<0$ there exist
$4$--dimensional \emph{\sgstatic}\footnote{We shall say that a
space-time is \emph{\sgstatic} if it contains a (strictly)
timelike Killing vector field which is orthogonal to the level
sets of a globally defined time function.}  solutions
$(\mcM,\lormet )$ of the vacuum Einstein equations with the
following properties:
\begin{enumerate}
\item \label{p1}$(\mcM,\lormet   )$ is diffeomorphic to $\R\times
\hyp$, for some $3$--dimensional spacelike Cauchy surface
$\hyp$, with the $\R$ factor corresponding to the action of the
isometry group. \item\label{p2} $(\hyp,\threeg)$, where $\threeg$
is the metric induced by $\lormet   $ on $\hyp$, is a complete
Riemannian manifold. \item\label{p3} $(\mcM,\lormet   )$ is
geodesically complete. \item \label{p4}All invariants of $\lormet
$ which are algebraically constructed using the curvature tensor
and its derivatives up to any finite order
 are uniformly bounded on $\mcM$.
 \item
\label{p5a} $(\mcM,\lormet   )$ admits a globally hyperbolic (in
the sense of manifolds with boundary\footnote{\label{fgh}We say
that a space-time with boundary is globally hyperbolic if it
contains a Cauchy surface; the latter are defined as hypersurfaces
which are intersected by every inextendible causal curve precisely
once.}) smooth conformal completion with a timelike $\Scri$.
\item\label{p5} $(\hyp,\threeg)$ has a $C^{\infty}$ conformal
compactification. \item \label{p6}The connected component of the
group of isometries of $(\mcM,\lormet )$ is exactly $\R$, with an
associated Killing vector $X$ being timelike throughout $\mcM$.
\item \label{p7}There exist no local solutions of the Killing
equation  other than the (globally defined) timelike Killing
vector field $X$.
\end{enumerate}
An example of a manifold satisfying points \ref{p1}-\ref{p5} above
is of course anti-de Sitter space-time. Clearly the anti-de Sitter
solution \emph{does not} satisfy points \ref{p6} and \ref{p7}.

One of the main results of this paper is a general existence
theorem producing a large class of space-times satisfying 1-8,
with prescribed data at conformal infinity.
\begin{theorem} \label{t1.1}
Any $C^{\infty}$ \sgstatic\ Lorentz metric $\gamma$ on ${\mathbb
R} \times S^{2}$, i.e.
$$\gamma = -\alpha^{2}dt^{2} + g_{S^{2}},$$
with $\alpha > 0$ and of non-negative scalar curvature on
${\mathbb R} \times S^{2}$, is the conformal infinity of a
complete, \sgstatic\ vacuum metric on ${\mathbb R}\times {\mathbb
R}^{3} = {\mathbb R}^{4}$, with cosmological constant $\Lambda =
-6$.
\end{theorem}

  Generic space-times in this result satisfy points 1-8 (if $\gamma$
  has another Killing vector, then $g$ also will; in particular the
construction also yields non-trivial static-axially symmetric vacuum
metrics). A related, but weaker, existence result is proved for space-times
with $\Lambda < 0$ and with conformal infinity of the topological form
${\mathbb R}\times T^{2}$.

 An existence result of this kind is also proved for space-times
with black holes of any given genus at the horizon and at
conformal infinity, (although again this result is not as general
as Theorem 1.1 in terms of the prescribed boundary data at
infinity,  see the discussion before \eq{e2.12}). In the black
hole context, Points 1-8 remain correct, with the following
obvious variations: Point~\ref{p1} remains correct if $\mcM$ is
replaced by its domain of outer communications. Point~\ref{p2}
remains correct if completeness is understood in the sense of
manifolds with (minimal) boundary. Finally, completeness holds for
geodesics starting in the domain of outer communication except
perhaps for those  which cross the future or past  event horizons.
The remaining points remain correct without modifications.

 Throughout this work we restrict attention to dimension
$n=4$. Similar but weaker results can be obtained in higher
dimensions; this will be discussed elsewhere. Let us simply
mention that arguments used below provide, e.g., existence of
vacuum $(n+1)$-dimensional \sgstatic\ metrics, with  the
associated boundary metrics lying in a neighborhood of
$-dt^2+g_{+1}$, where $g_{+1}$ is the round unit metric on
$S^{n-1}$ (this is the boundary metric arising from the
$(n+1)$-dimensional anti-de Sitter metric). Further, using
Kaluza-Klein reduction one can in particular
 obtain space-times as above, with or without
black hole regions, which satisfy the Einstein-Yang-Mills-dilaton
field equations; solutions belonging to this family have been
numerically constructed
in~\cite{Bjoraker:2000qd,Winstanley:2001bs}; see also
\cite{VanderBij:2001ia}.

It is clear that there should exist stationary \emph{and not
static} solutions as above, which can be constructed by solving an
asymptotic Dirichlet problem for the stationary Einstein equations
in a conformally compactifiable setting. We are planning to study
this question in a near future.

This paper is organised as follows: In
Section~\ref{S1} we review those results of \cite{mand1,mand2}
which are needed for our work here, and we give conditions under
which hypersurface-orthogonality descends from the conformal
boundary to the interior. In Section~\ref{S2} we prove our
existence and uniqueness results. In Section~\ref{Sglobal} we
justify the claims \ref{p1}-\ref{p7} made above.
%Appendix~\ref{Aext} contains results about extendibility of
%conformal isometries of $\partial M$ to isometries of $M$.
%\documentstyle\cite{mand2,1pt,amsfonts]{amsart}

%\newtheorem{theorem}{Theorem}[section]

\newtheorem{lemma}[defi]{\sc Lemma\rm}

\newtheorem{corollary}[defi]{\sc Corollary\rm}

\def\endproof{\qed\smallskip}
\def\blacksquare{\hbox to .60em{\vrule width .60em height .60em}}
\renewcommand{\qedsymbol}{\hfill \blacksquare \hspace *{4.5em}}

%\renewcommand{\theequation}{\thesection.\arabic{equation}}

%\begin{document}

\section{Isometry Extension}\label{Smie}\label{S1}

 Let $M$ be an oriented 4-manifold with compact boundary $\partial M$.
In this section, we work exclusively with Riemannian metrics. The results
obtained here will be applied to Lorentzian metrics in Section~\ref{S2}.

 Let $E = E_{AH}^{m,\alpha}$ denote the space of conformally compact or
asymptotically hyperbolic (AH) Einstein metrics on $M$; thus $g
\in E_{AH}^{m,\alpha}$ if $g$ is an Einstein metric on $M$,
$$Ric_{g} = -3g, $$
and $g$ has a $C^{m,\alpha}$ conformal compactification, $m \geq
2$, $\alpha  >  0$.  By this we mean that the metric $\rho^2 g$
extends by continuity to a $C^{m,\alpha}$ Riemannian metric on
$\bar M$, where $\rho$ is any function which vanishes precisely on
$\partial M$, and has nowhere vanishing gradient there. Let ${\cal
E} = {\cal E}_{AH}^{m,\alpha}$ be the moduli space of such
Einstein metrics, i.e. ${\cal E}_{AH}^{m,\alpha} =
E_{AH}^{m,\alpha}/{\cal D}_{1}$, where ${\cal D}_{1}$ is the group
of $C^{m+1,\alpha}$ diffeomorphisms of $\bar M$ which restrict to
the identity on $\partial M.$

 Let ${\cal C}  = {\cal C}^{m,\alpha}$ be the space of pointwise
conformal classes of $C^{m,\alpha}$ metrics on $\partial M.$ One
then has a natural boundary map
\begin{equation} \label{e1.1}
\Pi : {\cal E} \rightarrow  {\cal C}, \  \Pi[g] = [\gamma],
\end{equation}
which associates to each class $[g]$ its conformal infinity $[\gamma]$ on
$\partial M$.

 From \cite[Thm. A]{mand2}, ${\cal E}$ is either empty, or an infinite
dimensional Banach manifold, and $\Pi $ is a smooth Fredholm map
of Fredholm index 0. Thus, if non-empty, Im $\Pi$ is a ``variety''
of finite codimension in ${\cal C}$.
This result holds for any $m \geq 2$, in particular for $m = \infty$ or
for $m = \omega$, i.e. real-analytic.

 For $g \in {\cal E}$, let $K = K_{g}$ be the $L^{2}$ kernel of $D\Pi$,
i.e. the space of symmetric bilinear forms $h$ on $(M, g)$ which
are in $L^{2}(M, g)$ and satisfy $D\Pi(h) = 0$. At any $g \in
{\cal E}$ for which $K_{g} = 0$, the map $D\Pi$ is an isomorphism
and hence $\Pi({\cal E})$ contains an open neighborhood of $\gamma
= \Pi(g)$ in ${\cal C}$, cf.~\cite{GL,Biquard,Lee:fredholm}. Thus,
if ${\cal E}$ contains a regular point of $\Pi$, then Im$\Pi
\subset {\cal C}$ is a variety of codimension 0 in ${\cal C}$. It
is natural to conjecture that ${\cal E}$ always contains such a
regular point, and in fact such regular points are dense in ${\cal
E}$, although these issues are currently unresolved in general.

\medskip

  The starting point of the work in this section is the following
  result~\cite[Thm.~3.2]{mand2}:\footnote{It is unfortunate that Proposition~\ref{p 1.1} does not
necessarily hold for isometries which do not lie in the connected
component of the group, as such a property would have considerably
simplified some of the arguments below.}
\begin{proposition} \label{p 1.1}
 Let $g$ be an AH Einstein metric on a 4-manifold $M$,
$g \in {\cal E}_{AH}^{m,\alpha}$, $m \geq 4$, with conformal infinity
$[\gamma]$. Then any connected group $G$ of effective
conformal isometries of $(\partial M, \gamma)$ extends to a
connected group $G$ of effective isometries of $(M, g)$.
\end{proposition}

  In particular, if $(\partial M, \gamma) = (S^{3}, \gamma_{0})$ is the
round metric on $S^{3}$, then any AH Einstein filling metric has
an effective isometric $SO(3,1)$ action. The only such metric is
the hyperbolic metric $g_{-1}$ on the 4-ball $B^{4}$.\footnote{The
fact that the hyperbolic metric is the only Einstein metric on a
spin $4$-manifold with conformal infinity $(S^{3}, \gamma_{0})$
has already been established by rather different methods
in~\cite{AndDahl}.} In all other cases, by the well-known Obata
Theorem, (the conformal isometry group is equivalent to the
isometry group unless $\gamma $ is the round metric on $S^{3}),$
one may assume that $G$ is a group of isometries of $(\partial M,
\gamma)$.

  Thus, let $G = S^{1}$, and suppose that we have an isometric $S^{1}$
action on $(\partial M, \gamma)$. It follows that this action extends to an
isometric $S^{1}$ action on any AH Einstein filling metric $(M, g)$, with
associated Killing field $K$.

\medskip

  For an AH Einstein metric with an effective $S^{1}$ action on $M$,
there are in general three types of $S^{1}$ orbits:

\begin{itemize}
\item principal orbits, on which the action is free; \item
exceptional orbits, on which the action has a non-trivial but
finite isotropy group; \item fixed points.
\end{itemize}

  The fixed point set $F$ consists of a finite collection of smooth
surfaces $B = \cup B_{i}$ embedded in $M$, and a finite number of
isolated points $N = \cup n_{j}$; these are called bolts and nuts
respectively in the terminology of Gibbons-Hawking
\cite{GibbonsHawkingInstantons}. In general, the bolts might
intersect infinity, i.e. $\partial M$, and so be complete open
surfaces in $M$. This will be the case if the Killing field $K$
vanishes on circles in $\partial M$.  For simplicity, we rule out
zeros of $K$ at $\partial M$, and in fact we will only consider
free $S^{1}$ actions on $\partial M$.

  Let $\Sigma = M/S^{1}$ be the orbit space of the $S^{1}$ action, with
$\pi : M \rightarrow \Sigma$ the projection to the orbit space. By
\cite{Fintushel1,Fintushel2}, $\Sigma$ is topologically a
3-manifold with boundary, with $\partial \Sigma = B$. Of course
the nuts correspond to isolated points in $\Sigma$. Note that
$\Sigma$ also has a boundary at infinity $\partial_{\infty}\Sigma$
corresponding to the quotient $\partial M / S^{1}$. Let $E$ denote
the exceptional set, i.e. the union of the exceptional orbits in
$M$. Then the quotient $\Sigma_{E} = \pi(E)$ consists of a finite
number of smooth embedded curves $\sigma_{k}$ in $\Sigma$, along
which the isotropy is constant. Each curve $\sigma_{k}$ is either
a closed curve in the interior $\stackrel \circ \Sigma$, or is an
arc with distinct endpoints $\partial \sigma_{k} \subset \partial
\Sigma$. Let $P \subset M$ denote the union of the principal
orbits of the $S^{1}$ action. Then $P$ is open and dense in $M$
and $\Sigma_{P} = \pi(P)$ is open and dense in $\Sigma$.

  The metric $g$ on $M$ induces a metric $g_{\Sigma}$ on the orbit
space, for which the projection $\pi$ is a Riemannian submersion.
It follows that $g$ may be written in the form
\begin{equation} \label{e1.2}
g = u^{2}(d\phi  + \theta )^{2} + \pi^{*}g_{\Sigma},
\end{equation}
where $\theta $ is a connection 1-form, $u$ is the length of the
Killing field $K$ and $K = \partial /\partial\phi$. The parameter
$\phi $ parameterises a circle $S^{1}.$

  The nuts and bolts in $M$ are exactly the
locus where the Killing field $K$ vanishes, i.e. the 0-set of the
function $u$. It is well-known that sets of fixed points are
totally geodesic, cf. \cite{key485}, and hence the boundary
$\partial \Sigma$ is totally geodesic in $\Sigma$. The same
argument shows that the exceptional curves $\gamma_{k}$ are
geodesics in $(\Sigma, g_{\Sigma})$. The metric $g_{\Sigma}$ is
not smooth along each $\gamma_{k}$, but has normal cone
singularities, i.e. metrically the normal disc to $\gamma_{k}$ at
a point of $\gamma_{k}$ is of the form ${\Bbb R}^{2}/{\Bbb
Z}_{k'}$, where the cyclic group ${\Bbb Z}_{k'}$ acts in the usual
way by rotation on  ${\Bbb R}^{2}$; here of course the order $k'$
might depend on $k$.

\medskip

  We will use the following terminology. The $S^{1}$ action on $(M, g)$ is {\it \sgstatic}\ if
$(M, g)$ is globally a warped product of the form
\begin{equation} \label{e1.3}
M = S^{1}\times \Sigma, \ g = u^{2}d\phi^{2} + \pi^{*}g_{\Sigma},
\end{equation}
where $u: \Sigma \rightarrow {\Bbb R}$ is strictly positive and
$g_{\Sigma}$ is a complete metric on $\Sigma$, $\partial \Sigma =
\emptyset$. In this case, the $S^{1}$ action is just given by
rotations in the $S^{1}$ factor; there are no nuts, bolts or
exceptional orbits. The $S^{1}$ action is {\it \gstatic} if
\eq{e1.3} holds with $u = 0$ somewhere. In this case, the locus
$\{u = 0\}$ is given by bolts -- there are no nuts or exceptional
orbits. Next, the $S^{1}$ action is {\it topologically static} if
the $S^{1}$ bundle $S^{1} \rightarrow P \rightarrow \Sigma_{P}$ is
a trivial bundle, i.e. it admits a section. This is equivalent to
the existence of a cross-section of the $S^{1}$ fibration $P \cup
E \rightarrow \Sigma_{P\cup E}$. \footnote{$S$ will be called a
cross-section of a fibration if $S$ meets every fiber at least
once, with the intersection being transverse.} Finally, we  define
the $S^{1}$ action to be {\it locally static} if every point of
$(M, g)$ has a neighborhood isometric to a neighborhood of a point
with metric of the form \eq{e1.3}; this is equivalent to the usual
notion of static in terms of the existence of a hypersurface
orthogonal Killing field. It is easy to see that a locally static
action has no nuts.\footnote{\label{nonutsf}For a nut $n\in M$,
the geodesic $\varepsilon$-sphere $S_{n}(\varepsilon)$ is
invariant under the action, which is free. But
$S_{n}(\varepsilon)$ is topologically $S^{3}$ for sufficiently
small $\varepsilon$, which admits no free (even topologically)
static $S^{1}$ actions.}

 We shall use an obvious equivalent of the above for an $\R$ action
by isometries on a Lorentzian manifold $(\mcM,\lormet)$, with the
further restriction that the associated Killing vector field be
timelike almost everywhere.\
\medskip

 The main result of this section is the following:
\begin{theorem} \label{t 1.2}
  Let $(M, g)$ be as above, with $g \in {\cal E}^{m,\alpha}$, $m \geq 4$.
Suppose the free $S^{1}$ action at conformal infinity $(\partial M, \gamma )$
is \sgstatic, i.e.
\begin{equation} \label{e1.4}
\partial M = S^{1}\times V,
\end{equation}
and the $S^{1}$ action on $(M, g)$ is topologically static.

  Then the $S^{1}$ action on $(M, g)$ is \emph{locally static}, i.e.
$(M, g)$ is locally of the form \eq{e1.3}, (with $\{u = 0\} \neq \emptyset$
possibly).
\end{theorem}

\begin{remark} The result remains true if the hypothesis of
topological stability is replaced by the requirements that a)
$\zSigma$ is simply connected and b)  $\{u=0\}$ is connected, if
not empty. This can be established by an analysis of the equation
satisfied by the twist potential.
\end{remark}

 \proof The proof is
rather long, and will be broken into two steps.

\noindent {\bf Step I.}

  In this step, we set up the basic equations and obtain estimates on
the behavior near infinity.

 Let $\omega $ be the twist 1-form, given by
\begin{equation} \label{e1.5}
\omega  = \frac{1}{2}u^{3}*_{\Sigma}d\theta  =
\frac{1}{2}*_{M}(\kappa\wedge d\kappa),
\end{equation}
where $\kappa  = u^{2}(d\phi+\theta )=g(K,\cdot)$ is the 1-form
dual to the Killing field $K$, cf. \cite[(16.7),
(16.32)]{Exactsolutions}. Here, the first $*$ operator
$*_{\Sigma}$ is on $(\Sigma , g_{\Sigma}),$ while the second is on
$(M, g)$. The twist form is horizontal, i.e. $\omega(K) = 0$ and
is invariant under the $S^{1}$ action, so descends to a 1-form on
$\Sigma$.

  The condition that the $S^{1}$ action on $(M, g)$ is locally static is
equivalent to the condition that
$$\omega \equiv 0.$$
The Einstein equations imply that the 1-form $\omega $ is closed,
\begin{equation} \label{e1.6}
d\omega  = 0.
\end{equation}
It follows from \eq{e1.5} and \eq{e1.6} that, on the orbit space
$\Sigma$, one has
$$div(u^{-3}\omega ) = 0.$$
Hence, the divergence theorem applied to any bounded smooth domain
$U$ in $\Sigma$ implies
\begin{equation} \label{e1.7}
\int_{\partial U}u^{-3}\omega(\nu) = 0,
\end{equation}
where $\nu $ is the outward unit normal. In particular, choosing
$U$ to be a fixed connected neighborhood of any bolt $B_{i}$ and replacing
$U$ by $U\setminus B_{i}(\varepsilon)$, where
$B_{i}(\varepsilon)$ is the $\varepsilon$-tubular neighborhood of $B_{i}$,
it follows that
\begin{equation} \label{e1.8}
\int_{B_{i}}u^{-3}\omega(\nu)dA = \kappa_{i}
\end{equation}
is well-defined, where $\nu$ is a unit normal to $B_{i}$ in $\Sigma$.
Similarly, if $S_{j}(\varepsilon)$ is the
$\varepsilon$-sphere about the nut $n_{j}$, then the limit
\begin{equation} \label{e1.9}
\lim_{\varepsilon \rightarrow
0}\int_{S_{j}(\varepsilon)}u^{-3}\omega(\nu)dA = \kappa_{j},
\end{equation}
exists, (and is finite). These integrals give the nut charge of
each bolt or nut, in the terminology of
\cite[\S 5]{GibbonsHawkingInstantons}. Note also that since
$\omega \equiv 0$ on any bolt or nut, to leading order, $\omega
(\nu) \sim \pm |\omega|$ on $S_{j}(\varepsilon)$, where
$S_{j}(\varepsilon)$ is the boundary of the $\varepsilon$-tubular
neighborhood of the bolt or nut.

  Next we turn to the asymptotic behavior of the metric $g$.
Asymptotically, the metric $g = g_{\Sigma}$ on $\Sigma $ has the
form
\begin{equation} \label{e1.10}
g = r^{-2}dr^{2} + r^{2}\gamma + h,
\end{equation}
where $\gamma $ is the boundary metric on $\partial M,$ and $||h||
= O(r^{-2})$, w.r.t the metric $g$, cf.
\cite[\S 2]{mand2}. The function $u$ satisfies $u \sim
r$ for large $r$, i.e. the ratio $u/r$ is bounded away from $0$
and $\infty$. From \eq{e1.10}, observe that
\begin{equation} \label{e1.11}
|\nabla r| = |dr| = r.
\end{equation}

Let $\bar g = \rho^{2}g$ be a smooth, (at least $C^{4,\alpha}$),
compactification of $g$. Thus, $\rho  \sim  r^{-1}, u\rho  \sim 1$ and
\begin{equation} \label{e1.12}
\bar g = \rho^{2}u^{2}(d\phi  + \theta )^{2} + \rho^{2}g_{\Sigma}.
\end{equation}
To be concrete, choose $\rho$ to be the geodesic defining function
determined by a choice of boundary metric $\gamma \in [\gamma]$.
Since the Killing field preserves the class of geodesics
orthogonal to $\partial M$, $\rho$ is invariant under the $S^{1}$
action, and so descends to a defining function for
$\partial_{\infty}\Sigma$. In particular, $d\phi $ and $\theta $
are the same for $g$ and $\bar g$, while the norm of the Killing
field w.r.t. $\bar g$ is $\rho u$.

  Let $\bar \omega$ be the twist form, defined as in \eq{e1.5}, but w.r.t.
the $\bar g$ metric. Then $\bar \omega$ extends smoothly to $(\bar
M, \bar g)$, since the compactification is smooth. A direct
computation from \eq{e1.5} shows that, near $\partial M$,
$$2\bar \omega = (u\rho)^{4}\bar *_{M}[(d\phi +\theta )\wedge d\theta]  =
(u\rho)^{3}\bar *_{\Sigma} d\theta  =
2(u\rho)^{3}\rho^{-1}u^{-3}\omega.$$ Here, we use the fact that
$||d\phi||_{\bar g} = (u\rho)^{-1}$ and $\bar *_{\Sigma} =
\rho^{-1}*_{\Sigma}$ on 2-forms. Thus
\begin{equation}\label{e1.13}
\bar \omega = \rho^{2}\omega.
\end{equation}
In particular, one has
\begin{equation} \label{e1.14}
|\bar \omega|_{\bar g} = \rho|\omega|_{g}.
\end{equation}

  Since the action on the boundary is static, it follows that $\bar
\omega = 0$ on $\partial M$. Since the compactification is smooth,
$|\bar \omega|_{\bar g} = O(\rho)$, and hence \eq{e1.14} gives
$$|\omega| = O(1).$$

  We need to improve this estimate by order 1. This is done in the
following result.
\begin{lemma}\label{l 1.3}
The twist form $\omega = \rho^{-2} \bar \omega$ extends smoothly
to $\partial M$, with
\begin{equation} \label{e1.15}
|\omega| = O(\rho).
\end{equation}
\end{lemma}

\proof
 By \eq{e1.13} and \eq{e1.14}, it suffices to show that $\bar \omega = O(\rho^{2})$
near $\partial M$. Of course both $d\bar \omega$ and $\bar \omega$
are smooth up to $\partial M$.

  By \eq{e1.13}, we have
\begin{equation} \label{e1.16}
d\bar \omega = d(\ln \rho^{2}) \wedge \bar \omega.
\end{equation}

We claim first that if the action is static at $\partial M$, then
\begin{equation} \label{e1.17}
d \bar \omega = 0 \ {\rm at} \  \partial M.
\end{equation}
 To see this, by standard formulas, cf. \cite[p.10]{Heusler:book} for example, one has
in general
$$d\bar \omega = *(K \wedge \bar Ric(K)),$$
where $K$ is the Killing field w.r.t. $\bar g$, (and $K$ is
identified with the 1-form $\kappa$ by $\bar g$). Thus, the
vanishing -- or not -- of $d\bar \omega$ is governed by $\< \bar
Ric(K), X \>$, where $X$ is orthogonal to $K$. However, using the
formulas of~\cite[Lemma 1.2]{mand1} for instance, the ambient
Ricci curvature $\bar Ric$ at $\partial M$ is determined by the
intrinsic Ricci curvature of the boundary metric $\gamma$; in fact
for $X$, $Y$ tangent to $\partial M$,
$$\bar Ric(X,Y) = 2Ric_{\gamma}(X,Y) - \frac{1}{4}R_{\gamma}\cdot
\gamma(X,Y).$$ Now if the $S^{1}$ action on $\partial M$ is
static, then it is well-known that $Ric_{\gamma}(K, X) = 0$, for
any $X$ orthogonal to $K$ and tangent to $\partial M$. Similarly,
if $X = \bar \nabla \rho$, then $\bar Ric(K, X) = 0$,
by~\cite[Lemma 1.2]{mand1} again. This gives \eq{e1.17}. It
follows that $\rho^{-1}d\bar \omega$ extends smoothly to $\partial
M$.

  By \eq{e1.16}, this means that $\rho^{-2}\bar \omega(X)$ extends smoothly
up to $\partial M$, for any smooth vector field $X$ orthogonal to
$\bar \nabla \rho$. Thus, it remains to understand the behavior of
$\bar \omega(\bar \nabla \rho)$.
  By \eq{e1.6}, one has
$$\bar \omega(\bar \nabla \rho) = \frac{1}{2}\rho\; u\; d\kappa
(T_{1},T_{2}) = -\frac{1}{4}\rho\, u\, \bar g( K, [T_{1}, T_{2}]
)\;,$$ where $K/|K|$, $T_{1}$ and $T_{2}$ are $\bar
g$--orthonormal vector fields orthogonal to $\bar \nabla \rho$.
Hence, at $\partial M$, $T_{1}$ and $T_{2}$ are tangent to the
surface $V$ in \eq{e1.4}. For further calculations it will be
convenient to assume that $\bar
\nabla_{T_{i}}T_{j}=0$ % ={}^\gamma \nabla_{T_{i}}T_{j} = 0$
at the given (arbitrary) point in $V$.
%, where the covariant
%derivative ${}^\gamma \nabla$ is taken in $(V, \gamma)$.
This is compatible with the requirement of tangency of the $T_i$'s
to $V$ because $\partial M$ is totally geodesic in $\bar M$, with
respect to the $\bar g$ metric. Since the $S^{1}$ action is static
at $\partial M$, one has
\begin{equation} \label{e1.18}
\bar \omega(\bar \nabla \rho) = 0 \ {\rm at} \ \partial M,
\end{equation}
compare with \eq{e1.14}. Thus, it suffices to prove that
$$\bar \nabla \rho(\bar \omega(\bar \nabla \rho)) = 0 \ {\rm at} \
\partial M.$$
This is a direct computation: set $\beta = -\frac{1}{4}\rho u$
(which has a smooth limit on $\partial M$, still denoted by the
same symbol) so that, using \eq{e1.18} one has
$$\bar \nabla \rho(\bar \omega(\bar \nabla \rho))|_{\partial M} =
\beta \bar g( \bar \nabla_{\bar \nabla \rho}K, [T_{1},T_{2}] ) +
\beta \bar g( K, \bar \nabla_{\bar \nabla \rho}\bar
\nabla_{T_{1}}T_{2} - \bar \nabla_{\bar \nabla \rho} \bar
\nabla_{T_{2}}T_{1} \rangle.$$ For the first term, since $K$ is
conformal Killing on $\partial M$, one has $\bar g( \bar
\nabla_{\bar \nabla \rho}K, [T_{1},T_{2}] )= - \bar g( \bar
\nabla_{[T_{1},T_{2}]}K, \bar \nabla \rho )+ \lambda \bar g( \bar
\nabla \rho, [T_{1},T_{2}] )$, both of which vanish since the
$2^{\rm nd}$ fundamental form $A = 0$ at $\partial M$.

  To calculate the second term, observe first that for any
horizontal field $T$ on $\partial M$, i.e. $T$ tangent to the $V$
factor in \eq{e1.4}, $\bar \nabla_{\bar \nabla \rho}T$ is also
horizontal. To see this, using the fact that $K$ is conformal
Killing on $\partial M$, one has $\bar g( \bar \nabla_{\bar \nabla
\rho}T, K )= - \bar g( \bar \nabla_{\bar \nabla \rho}K, T )= \bar
g( \bar \nabla_{T}K, \bar \nabla \rho )= 0$, since again $A = 0$.
Now write $$\bar g( K, \bar \nabla_{\bar \nabla \rho}\bar
\nabla_{T_{1}}T_{2} )= \bar g( K, \bar \nabla_{T_{1}} \bar
\nabla_{\bar \nabla \rho}T_{2} )+ \bar g( \bar R(\bar \nabla \rho,
T_{1})T_{2}, K )+ \bar g( \bar \nabla_{[\bar \nabla
\rho,T_{1}]}T_{2}, K \rangle\;.$$  For the curvature term, the
Gauss-Codazzi equations give $\bar g( \bar R(\bar \nabla
\rho,T_{1})T_{2}, K )= dA(T_{1},T_{2},K)$, which vanishes since $A
= 0$ at $\partial M$. Further, again since $A = 0$, $[\bar \nabla
\rho, T]$ is tangent to $V$, and hence $\bar g( \bar \nabla_{[\bar
\nabla \rho,T_{1}]}T_{2}, K )= 0$ by \eq{e1.18}. Finally, for the
first term above, since $\bar \nabla_{\bar \nabla \rho}T_{2}$ is
horizontal, so is $\bar \nabla_{T_{1}} \bar \nabla_{\bar \nabla
\rho}T_{2}$, and hence the first term also vanishes.

  This completes the proof. In fact, with further computation, it is not
difficult to show that $\bar \omega(\bar \nabla \rho)$ vanishes to
order 2 at $\partial M$.
{\endproof}

\medskip
\noindent {\bf Step II.}

 The remainder of the proof is an adaptation of a result of Heusler,
\cite[Thm.8.2]{Heusler:book}, on stationary asymptotically flat
Lorentz metrics to the current situation. The basic tool is the
following Lemma, proved in the Lorentzian setting in
\cite[(8.4)]{Heusler:book}. The proof in the Riemannian setting is
essentially the same, but for clarity we give a complete proof.

\begin{lemma} \label{l 1.4}
 Let $(M, g)$ be a Riemannian $4$-manifold, with Killing field $K$, with
$u^{2} = |K|^{2}$ and with twist 1-form $\omega$. Then
\begin{equation} \label{e1.19}
d(\omega\wedge\frac{\kappa}{u^{2}}) = d\omega
\wedge\frac{\kappa}{u^{2}} -  2\frac{|\omega|^{2}}{u^{4}}*\kappa ,
\end{equation}
where $\kappa $ is the 1-form dual to $K$ by the metric.
\end{lemma}

\proof
 Clearly it suffices to show that
\begin{equation} \label{e1.20}
\omega\wedge d(\frac{\kappa}{u^{2}}) =
2\frac{|\omega|^{2}}{u^{4}}*\kappa .
\end{equation}
To do this, we first claim that
\begin{equation} \label{e1.21}
d(\frac{\kappa}{u^{2}}) = 2*(\kappa\wedge\frac{\omega}{u^{4}}).
\end{equation}

 To verify \eq{e1.21}, a simple computation shows that for any 1-form
$\alpha$ on $M^{4},$ one has
\begin{equation} \label{e1.22}
\iota_{X}(*\alpha ) = *(\alpha\wedge\xi ),
\end{equation}
where $\xi $ is the 1-form dual to the vector field $X$. Applying
this to $X = K$ and $\alpha  = \omega $ gives
$$*(\kappa\wedge\omega ) = -\iota_{K}(*\omega ) = +\frac{1}{2}\iota_{K}(\kappa\wedge d\kappa ), $$
where the last inequality follows from the definition of $\omega $
and the fact that $*^{2} = - 1$ for 3-forms on a 4-manifold.
Interior multiplication is a skew-derivation, and so
$\iota_{K}(\kappa\wedge d\kappa ) = \iota_{K}(\kappa )d\kappa  -
\kappa\wedge\iota_{K}(d\kappa ).$ One has $\iota_{K}(\kappa ) =
|K|^{2} = u^{2},$ while, since ${\cal L}_{K}\kappa = 0$,
 $\iota_{K}(d\kappa ) = - d\iota_{K}(\kappa ) = - du^{2}.$ This gives
$$2*(\kappa\wedge\omega ) = u^{2}d\kappa  + \kappa\wedge du^{2}, $$
or equivalently,
$$u^{-2}d\kappa  = -  u^{-4}\kappa\wedge du^{2} + 2u^{-4}*(\kappa\wedge\omega ) . $$
But $d(u^{-2}\kappa ) = - u^{-4}du^{2}\wedge\kappa  +
u^{-2}d\kappa ,$ which then gives \eq{e1.21}.

 From \eq{e1.21}, it follows that
$$\omega\wedge d(\frac{\kappa}{u^{2}}) = 2u^{-4}\omega\wedge *(\kappa\wedge\omega ) =
-2u^{-4}\omega\wedge\iota_{K}(*\omega ), $$
where the second equality follows from \eq{e1.22}. But
$\iota_{K}(\omega\wedge *\omega ) = \iota_{K}(\omega )\wedge
*\omega  -  \omega\wedge\iota_{K}(*\omega ).$ Since
$\iota_{K}\omega = 0$, ($\omega $ is horizontal while $K$ is
vertical), one obtains
$$\omega\wedge d(\frac{\kappa}{u^{2}}) = 2u^{-4}\iota_{K}(\omega\wedge *\omega ) = 2u^{-4}
|\omega|^{2}\iota_{K}d\mathit{vol}_{M}, $$ which gives \eq{e1.20}.
{\endproof}

 We apply this result to the AH Einstein manifold $(M, g)$. By \eq{e1.6},
$d\omega  = 0$, and so \eq{e1.19} implies
\begin{equation} \label{e1.23}
2\frac{|\omega|^{2}}{u^{4}}*\kappa  = - d(\omega\wedge\frac{\kappa}{u^{2}}) .
\end{equation}
Now by assumption, the $S^{1}$ action on $(M, g)$ is topologically
static. Hence, there exists a cross-section $\iota: \Sigma_{P\cup
E} \rightarrow (P \cup E) \subset M$, such that $\pi \circ \iota =
id$. Let $S$ = Im$\iota$. Thus $S$ is an embedded hypersurface in
$M$, which may be perturbed slightly if necessary to obtain a
smooth hypersurface $S$, transverse to the $S^{1}$ orbits. In
addition, we may assume, without loss of generality, that $S$ is a
3-manifold with boundary $\partial S$, (possibly empty). The
boundary consists of bolts, together possibly with a finite
collection of points corresponding to the nuts. (Hence, these are
not actually in $\partial S$, but just distinguished points in
$S$. The metric $\bg$ might have a singularity there, which is
sufficiently mild to be irrelevant for our purposes here.).
Moreover, $S$ has a boundary at infinity,
\begin{equation} \label{e1.24}
\partial_{\infty}S = V,
\end{equation}
for $V$ as in \eq{e1.4}, which may be assumed to be smooth, (in
the given compactification $(\bar M, \bar g)$). Thus, $(S, g_S)$,
where $g_S = g|_{S}$ is a complete Riemannian 3-manifold with
boundary. The fact that $\partial_{\infty}S = V$ smoothly implies
that we may assume the 3-manifold $S$ becomes orthogonal to the
$S^{1}$ orbits on approach to $\partial M,$ i.e.
\begin{equation} \label{e1.25}
\langle K/|K|, e \rangle (x) \rightarrow  0, \ \ {\rm as} \ \ x \rightarrow
\partial M,
\end{equation}
where $e$ is any unit tangent vector to $S$ at $x$.

 Now integrate \eq{e1.23} over $S$ and apply Stokes theorem to obtain
\begin{equation} \label{e1.26}
2\int_{S}\frac{|\omega|^{2}}{u^{4}}*\kappa  = -\int_{\partial
S}\omega\wedge\frac{\kappa}{u^{2}} .
\end{equation}
On $S$, the 3-form $*\kappa  = \alpha|K|dvol_{S},$ where $\alpha $
is the sine of the angle between $TS$ and $K$, (at any given $x\in
S$). Since $S$ is everywhere transverse to $K$, $\alpha  > 0$
everywhere, (or $< 0$ everywhere, depending on the choice of
orientation). Hence, the integrand on the left in \eq{e1.26} does
not change sign.

 The boundary integral on the right in \eq{e1.26} has two parts -- the integral
over the boundary at infinity $\partial_{\infty}S,$ and over the
nuts and bolts $\partial S$; these will be treated separately.

 First, for the integral at infinity, using the fact that $S$ approximates
$\Sigma $ near infinity, by prior estimates, we have
$$|\omega| = O(r^{-1}), \ |\kappa| = O(r), \ u^{2} = O(r^{2}), \ \mbox{\rm and} \
\mbox{\em area} S(r) = O(r^{2}). $$ It follows that the boundary
integrand at infinity in \eq{e1.26} is bounded. Further,
\eq{e1.25} implies that for any unit tangent vector in $TS$,
$(\kappa /|\kappa|)(e) \rightarrow 0$ at infinity. Hence the
boundary integral at infinity vanishes.

 For the boundary integral over the bolts and nuts, as noted in \eq{e1.8} and
\eq{e1.9}, the integral of $u^{-3}|\omega|$ over bolts and nuts in
the orbit space $\Sigma $ is finite, i.e. the integral over the
boundary of an $\varepsilon$-tubular neighborhood tends to a
finite limit as $\varepsilon  \rightarrow 0$. Since the volume of
such a boundary approximation in the manifold $M$ is small
compared with that in $\Sigma$, and since in addition $|\kappa|$
is bounded, ($|\kappa| \rightarrow 0$ in fact), it is clear that
the boundary integral over bolts and nuts also vanishes.

 It follows then from \eq{e1.26} that $\omega \equiv 0$ on $S$. Since
$\omega $ is also invariant under the $S^{1}$ action, it follows
that $\omega  \equiv 0$ on $M$. This implies that $(M, g)$ is
everywhere locally static.

%\ptc{comment about global staticity
%removed}
% Together with the assumption that $(M,
%g)$ is topologically static, and the action at $\partial M$ is
%globally static,\ma{comment added} it follows that $(M, g)$ is
%\gstatic.
{\endproof}

  We complete this section with the following topological result, needed
for the work in the next section.\footnote{Useful discussions with
A.Zeghib concerning  this Lemma are acknowledged.}

%\begin{lemma} \label{l2.5old}\ptc{old version; my suggestion for a new
%version below} Let $M$ be a compact
%4-manifold with boundary. Suppose $M$ has a topologically static
%$S^{1}$ action $\Phi_{0}: S^{1} \rightarrow {\rm Diff}(M)$. If
%$\Phi_{1}: S^{1} \rightarrow {\rm Diff}(M)$ is any other $S^{1}$
%action obtained by deformation from $\Phi_{0}$, i.e. there is a
%continuous curve $\Phi_{t}$ of $S^{1}$ actions joining $\Phi_{0}$
%and $\Phi_{1}$, then $\Phi_{1}$ is also topologically static.
%\end{lemma}

%\ptc{End of old version; new version starts here; note the new
%statement and proof. You can revert to the old proof if you wish.}

\begin{lemma} \label{l2.5} Let $M$ be a compact
4-manifold with boundary. Suppose $M$ has a topologically static
$S^{1}$ action $\Phi_{0}: S^{1} \rightarrow {\rm Diff}(M)$ without
exceptional orbits. If $\Phi_{1}: S^{1} \rightarrow {\rm Diff}(M)$
is any other $S^{1}$ action obtained by deformation from
$\Phi_{0}$, i.e. there is a continuous curve $\Phi_{t}$ of $S^{1}$
actions joining $\Phi_{0}$ and $\Phi_{1}$, then $\Phi_{1}$ is also
topologically static without exceptional orbits.
\end{lemma}

\noindent \proof This follows from the topological classification
theorem in~\cite{Fintushel1}. Namely, smooth $S^{1}$ actions on
$M$ are uniquely determined, (up to smooth conjugacy), by discrete
data associated to the orbit space $\Sigma$. But such data do not
vary under continuous deformation of the action.

  More directly, the structure of the orbits near fixed points,
together with elementary topological considerations, shows that
both the existence of the set of fixed points, its topology, as
well as the topological properties of the orbit space near this
set are stable under continuous deformations of the metric. It is
then easy to reduce the problem to one where $\Phi_{0}$ has no
fixed points. The properties just mentioned imply then that none
of the $\Phi_t$'s has fixed points. The openness of the property
``having a slice" is obvious. To show that this property is
closed, consider a sequence of actions $\Phi_{t_i}$ with $t_i$
converging to $t_*$. For each $t_i$ the orbit space is a trivial
principal bundle, hence has vanishing characteristic class.  By
continuity, the $t_*$--orbit
 space will therefore also be a principal bundle with vanishing
 characteristic class, consequently  a trivial principal bundle,
 as desired.
{\endproof}

  More general results of this kind may well hold; however Lemma \ref{l2.5}
suffices for our needs. Note that it is easy to construct stationary and
not static $S^{1}$ actions on any $4$-manifold which admits
topologically static $S^{1}$ actions.

\section{Existence and Uniqueness}
\label{Sexu}\label{S2}

  In this section, we apply the results of Section~\ref{S1} to construct large
classes of non-trivial static AdS vacuum space-times. In addition, several
uniqueness results are proved.

  Throughout this section, it is assumed that $(\partial M, \gamma)$ has a
free \gstatic\ isometric $S^{1}$ action, i.e. $(\partial M,
\gamma)$ is \sgstatic. Hence $\partial M$ is topologically of the
form
\begin{equation} \label{e2.0}
\partial M = S^{1}\times V ,
\end{equation}
where $V$ is a closed surface. We work first in the Riemannian setting, and
pass later to Lorentzian metrics. The metric $\gamma $ on $\partial M$
is of the form
\begin{equation} \label{e2.1}
\gamma  = \alpha^{2}d\phi^{2}+ g_{V},
\end{equation}
with $\alpha > 0$ the length of the $S^{1}$ orbits. Modulo these
restrictions, the metric $\gamma$ may a priori be arbitrary, i.e.
$\alpha$ and $g_{V}$ are arbitrary. It is worthwhile to make a
brief dimension count here. The metric $\gamma$ corresponds to 2
free functions on $V$, after dividing out by diffeomorphisms of
$V$; namely $\alpha $ and a conformal factor for $g_{V}$. Thus,
$[\gamma]$, after dividing by diffeomorphisms, corresponds to 1
free function on $V$.

 Let ${\cal C}_{S} = {\cal C}_{S}(\partial M) \subset {\cal C}(\partial M)$
be the space of pointwise conformal classes of such static
boundary metrics. One can choose representatives of the conformal
class in several natural ways; by requiring that $g_{V}$ is of
constant curvature, by requiring $\alpha  = 1$, or by requiring
$\gamma$ is a Yamabe metric on $\partial M$. Of course generally
these normalisation are all distinct. Observe that the space
${\cal C}_{S}$ is connected.

\medskip

 Let $M$ be any compact, connected oriented 4-manifold, with boundary
equal to $\partial M$, and let ${\cal E}_{S} = {\cal E}_{S}(M)$ be the
moduli space of AH Einstein metrics on $M$, with boundary metrics in
${\cal C}_{S}$, i.e.
\begin{equation} \label{e2.2}
\Pi^{-1}({\cal C}_{S}) = {\cal E}_{S},
\end{equation}
for $\Pi$ the boundary map as in \eq{e1.1}.

By Proposition \ref{p 1.1}, any AH Einstein filling metric $(M, g)
\in {\cal E}_{S}$ of $(\partial M, \gamma)$, (if it exists), has
an effective isometric $S^{1}$ action. If this action is
topologically static on $M$, then Theorem \ref{t 1.2} implies that
the $S^{1}$ action on $(M, g)$ is static. Finally, Lemma
\ref{l2.5} shows that the condition that the $S^{1}$ action is
topologically static without exceptional orbits is stable, i.e.
preserved under deformations. (This holds actually regardless of
any equations.) It is clear that the isometric $S^{1}$
action induced on any given $g \in {\cal E}_{S}$ varies
continuously with any continuous variation of $g$ in ${\cal
E}_{S}$. It follows that on any given component ${\cal E}_{S}^{c}$
of ${\cal E}_{S}$, either every metric $g\in {\cal E}_{S}^{c}$ is
\gstatic\ without exceptional orbits, or no $g\in {\cal
E}_{S}^{c}$ has this property. Similarly the topology of the set
of fixed points of the action is constant on any connected
component of ${\cal E}^c_S$.

  In the following, we always work on a \gstatic\ component
$g\in {\cal E}_{S}^{c}$ of ${\cal E}_{S}$, and for simplicity write
${\cal E}_{S}^{c}$ as ${\cal E}_{S}$, when there is no danger of confusion.
The metric $g \in {\cal E}_{S}$ then has the form
\begin{equation} \label{e2.3}
g = u^{2}d\phi^{2} + g_{\Sigma},
\end{equation}
of a warped product on $M = \Sigma \times_{u}S^{1}$, modulo
diffeomorphisms in ${\cal D}_{1}$, i.e. equal to the identity on
$\partial M$. In particular, there are no exceptional orbits or
nuts$^{\mbox{\scriptsize \ref{nonutsf}}}$, only bolts, (if $u$
vanishes somewhere). The $S^{1}$ action is given by rotation about
the $S^{1}$ factor and the 3-manifold $\Sigma$ has boundary at
infinity given by $\partial_{\infty}\Sigma  = V$. The boundary map
\eq{e1.1} thus naturally restricts to the boundary map
$$\Pi_{S}: {\cal E}_{S} \rightarrow  {\cal C}_{S},$$
on globally static components ${\cal E}_{S}^{c} \subset {\cal E}_{S}$.
  Both spaces ${\cal C}_{S}$ and ${\cal E}_{S}$ are smooth infinite
dimensional Banach manifolds, (again assuming ${\cal E}_{S} \neq \emptyset$).
In fact both are closed submanifolds of ${\cal C}$ and ${\cal E}$
respectively. The restricted boundary map $\Pi_{S}$ is again a smooth
Fredholm map of Fredholm index 0.

\medskip

 Next, let ${\cal C}^{o}$ be the space of conformal classes of metrics on
$\partial M$ containing non-flat representative of non-negative
scalar curvature. Similarly, let ${\cal C}_{S}^{o}$ be the
corresponding space of \sgstatic\ boundary metrics. While it is
unknown if ${\cal C}^{o}$ is connected or not, the space ${\cal
C}_{S}^{o}$ is connected, by the uniformization theorem for
surfaces. As in \eq{e2.2}, define
$${\cal E}^{o} = \Pi^{-1}{\cal C}^{o}, \  {\cal E}_{S}^{o} = \Pi^{-1}({\cal C}_{S}^{o}).$$
By \cite[Thm.B]{mand2}, the boundary maps $\Pi$ and $\Pi_{S}$,
restricted to ${\cal E}^{o}$ and ${\cal E}_{S}^{o}$ respectively,
are proper maps, provided the map on homology induced by the
inclusion, $H_{2}(\partial M, {\Bbb R}) \rightarrow H_{2}(M, {\Bbb
R})$ is surjective. (This will be the case in all situations
discussed below). In particular, there are only finitely many
components of ${\cal E}^{o}$ and ${\cal E}_{S}^{o}$. As above, in
the following, we will always work in a fixed \gstatic\ component
of ${\cal E}_{S}^{o}$.

  Moreover, the maps $\Pi$ and $\Pi_{S}$ have a well-defined degree on
${\cal E}^{o}$ and ${\cal E}_{S}^{o}$, with
\begin{equation}\label{e2.4}
deg\Pi = deg \Pi_{S} \in {\Bbb Z}.
\end{equation}
The degree is computed by counting, with suitable signs, the number of points
in the inverse image of a regular value of $\Pi$, resp. $\Pi_{S}$. (Of course
the degree may change on different components of ${\cal E}^{o}$).

\medskip

  All of this discussion now carries over to the Lorentzian setting,
by changing $d\phi^{2}$ to $- dt^{2}$ and unwrapping the $S^{1}$
to ${\Bbb R}$. The boundary metric then has
the Lorentz form
\begin{equation} \label{e2.5}
\gamma  = \gamma^{L} = -\alpha^{2}dt^{2} + g_{V}.
\end{equation}
Further, it is clear that each AH (Riemannian) Einstein metric
\eq{e2.3} corresponds uniquely to a Lorentz metric of the form
\begin{equation} \label{e2.6}
\lormet = - u^{2}dt^{2} + g_{\Sigma},
\end{equation}
and vice versa. The metric $\lormet$ is a solution to the vacuum Einstein
equations with $\Lambda  = -6$. The bolts $B = \partial\Sigma  = \{u
= 0\}$ correspond to a horizon in the resulting space-time in the
usual way.

\medskip

 We now separate the discussion into distinct cases, according to
the genus of $V$.

\subsection{$V = S^{2}.$}

 Then $\partial M = {\Bbb R}\times S^{2},$ or $S^{1}\times S^{2}$ in
the Riemannian setting. Working first in the Riemannian setting,
suppose that
\begin{equation} \label{e2.7}
M = S^{1}\times {\Bbb R}^{3}.
\end{equation}
The manifold $M$ carries a \sgstatic\ AH Einstein metric, namely the
hyperbolic metric $g_{0}$ on $H^{4}(-1)/{\Bbb Z}$, where the action of
${\Bbb Z}$ is generated by a hyperbolic translation along a geodesic.
The $S^{1}$ action on $M$ and $\partial M$ are given by rotation in the
$S^{1}$ factor in (3.8). Let ${\cal E}$, ${\cal E}_{S}$ and
${\cal E}_{S}^{o}$ be the corresponding components containing $g_{0}$.
Hence, by the discussion above, ${\cal E}_{S}$ and ${\cal E}_{S}^{o}$
consist of \gstatic\ AH Einstein metrics on $M$.

\medskip

{\bf Existence:}
 By \cite[\S 7]{mand2}, for $M$ as in \eq{e2.7},
\begin{equation} \label{e2.8}
deg \Pi = deg \Pi_{S} = 1.
\end{equation}
Since $deg \Pi \neq 0$ implies $\Pi$ is surjective, it follows that any
$C^{m, \beta}$ boundary metric of non-negative scalar curvature is the
boundary metric of an AH Einstein metric on $M$, $m \geq 4$. More to the
point here, $deg \Pi_{S} \neq 0$ implies that $\Pi_{S}$ is surjective onto
${\cal C}_{S}^{o}$, and hence  any class in ${\cal C}_{S}^{o}$ is the
conformal infinity of a \gstatic\ AH Einstein metric on $M$. Observe that
$$u >  0 \ {\rm on} \ M, $$
so that $(M, g)$ is in fact \sgstatic. This follows by a simple
topological argument.  Namely suppose $u (p)= 0$ somewhere in $M$,
and for $\exp{(i\alpha)}\in S^1$, $q\in S^1\times\R^3$ let
$\phi(\alpha,q)$ denote the action of $S^1$ on $S^1\times\R^3$. By
construction the orbits $\alpha\to \phi(\alpha,q)$ are
homotopically non-trivial circles for all $q$ near the conformal
boundary. Let $q$ be such a point and for $s\in [0,1]$ let
$\gamma(s)$ be any continuous path from $p$ to $q$. Then
$\phi(\alpha,\gamma(s))$ provides a continuous deformation of
$\{\phi(\alpha,\gamma(1)=q)\}_{\alpha\in[0,2\pi]}$ to
$\{p\}=\{\phi(\alpha,\gamma(0)=p)\}_{\alpha\in[0,2\pi]}$, which is
not possible.

 Passing to the Lorentzian setting, the following corollary -- which
implies Theorem 1.1 -- is now immediate.
\begin{corollary} \label{c 2.1.}
Any $C^{m,\beta}$, $m \geq 4$, \gstatic\ boundary metric
$$\gamma  = -\alpha^{2}dt^{2} + g_{S^{2}}, $$
with $\alpha > 0$ and of non-negative scalar curvature on ${\Bbb
R}\times S^{2}$, is the conformal infinity of a complete,
\sgstatic\ vacuum AdS metric on ${\Bbb R}\times {\Bbb R}^{3} =
{\Bbb R}^{4}.$
\end{corollary}
{\endproof}

{\bf Uniqueness:} It is not known in general whether there is a
unique static Riemannian, (or Lorentzian), Einstein metric on
$(S^{1}\times {\Bbb R}^{3}, g)$ filling in a given boundary metric
$\gamma \in {\cal C}^{o}$. The boundary map $\Pi_{S}$ is of degree
1. The Sard-Smale theorem~\cite{Smale} implies that the regular
values of $\Pi_{S}$ are generic, and in particular dense, in
${\cal C}_{S}^{o}$. Hence a generic boundary metric in ${\cal
C}_{S}$ has an {\it odd} number of Einstein filling metrics in the
component ${\cal E}_{S}^{o}$. There are at present however no good
arguments in favor of uniqueness in general.

 On the other hand, in the presence of more symmetry, one can obtain
uniqueness; a similar result has been already established
in~\cite{ChHerzlich} (extending a previous argument of~\cite{BGH})
by completely different methods:

\begin{proposition} \label{p 2.2.}
  The standard anti-de Sitter metric $g_{o}$ on $\R\times {\Bbb R}^{3}$ is the
unique  \gstatic\ AdS vacuum metric on $\R\times {\Bbb R}^{3}$
which admits a $C^{2}$ (or $L^{2,p}, p >  4$) conformal
completion, with a conformally compactifiable smooth acausal
hypersurface orthogonal to the static Killing field, and which has
boundary metric $\gamma_{0} = - dt^{2} + g_{+1},$ where $g_{+1}$ is the
round metric on $S^{2}$.

%\ptc{"conformal compactification" before was wrong, the
%$\R$ factor was not compactified (unless we agree to do that, but
%then this should be explicitly said) --- changed to "conformal
%completion"; condition of compactifiabLE acausal hypersurfaces
%added}

\end{proposition}

 \proof
 By the correspondence Lorentzian -- Riemannian discussed above, it
suffices to prove the Riemannian analogue of this. This has
already been done in \cite[\S 7]{mand2}, using the fact that the
large isometry group $S^{1}\times SU(2)$ of the Riemannian
boundary metric extends to the same isometry group of any AH
Einstein metric, together with the classification of Einstein
metrics with such a large isometry group on the given manifold. {\endproof}

\medskip

 Next, for $V = S^{2}$ as before, assume that
\begin{equation} \label{e2.9}
M = {\Bbb R}^{2}\times S^{2}
\end{equation}
in the Riemannian setting. The manifold $M$ carries a natural
\gstatic\ AH Einstein metric, namely the Riemannian (or Euclidean)
Schwarzschild metric
\begin{equation} \label{e2.10}
g_{\mathrm{Schw}} = u^{2}d\phi^{2} + u^{-2}dr^{2} + r^{2}g_{S^{2}(1)},
\end{equation}
where $u^{2} = 1+r^{2} - \frac{2m}{r}$, $m > 0$. The period of $\phi$
depends on $m$, see the discussion below following (3.13). The boundary
metric $\gamma_{0}$ of $g_{\mathrm{Schw}}$ is clearly in ${\cal C}_{S}^{o}$
while $g_{\mathrm{Schw}} \in {\cal E}_{S}^{o}$, where the $S^{1}$ action is
given by rotation in the $\phi$-circles. Hence, as above, the full component
${\cal E}_{S}^{o}$ containing $g_{\mathrm{Schw}}$ consists of \gstatic\
AH Einstein metrics on $M$.

  For simple topological reasons as before, any metric in ${\cal E}_{S}^{o}$
has a unique bolt $B$, $B = S^{2}$.

\medskip

{\bf Existence:}
 By \cite[\S 7]{mand2},
$$deg \Pi  = 0 $$
and $\Pi $ is {\it not} onto ${\cal C}^{o}$. Similarly, $\Pi_{S}$
is not onto ${\cal C}_{S}^{o}$. In addition, uniqueness also
fails, in that for any regular value of $\gamma \in {\cal C}^{o}$
there are an {\it even} number, (possibly 0),  of Einstein filling
metrics.

  Nevertheless, the moduli spaces ${\cal E}$ and ${\cal E}_{S}$
are non-empty, since the Schwarzschild metrics \eq{e2.10} are in
${\cal E}_{S}$. Further, for a generic value of the mass $m$, the
Schwarzschild metrics are regular points of $\Pi$,
cf.~\cite{Hawking:1983dh} and hence $\Pi({\cal E})$ and $\Pi({\cal
E}_{S})$ are varieties of codimension 0, (containing open sets),
in ${\cal C}$ and ${\cal C}_{S}$ respectively.

  To any metric $g \in {\cal E}_{S}$ is associated a \gstatic\
Lorentzian-Einstein metric on the manifold
\begin{equation} \label{e2.11}
\mcM= {\Bbb R}\times ({\Bbb R}^{+}\times S^{2}),
\end{equation}
with metric of the form \eq{e2.6}. The usual extensions of the
space-time lead to smooth horizons, topologically ${\Bbb R}\times
S^{2}$, within $\mcM$.

  This gives the existence of a large, infinite dimensional, family of
smooth, conformally compact vacuum AdS black hole metrics on
$\mcM$ as in \eq{e2.11}. This is in marked contrast to the black
hole uniqueness theorem in case $\Lambda = 0$. Of course the main
example is the AdS Schwarzschild metric, (or spherical Kottler
metric),
\begin{equation} \label{e2.12}
g_{\mathrm{Schw}}  = - u^{2}dt^{2} + u^{-2}dr^{2} + r^{2}g_{S^{2}(1)}.
\end{equation}

  We point out however an interesting difference in the behavior of the
boundary maps $\Pi$ in the Riemannian and Lorentzian settings.
Thus, note that as $m$ varies, the conformal class of the boundary
metric of the Riemannian $g_{\mathrm{Schw}}$ in \eq{e2.10} varies;
the boundary metric is of the form
$$d\phi^{2} + g_{S^{2}(1)},$$
for $\phi \in [0, \beta]$, where $\beta  =  4\pi
r_{+}/(1+3r_{+}^{2})$, and $r_{+}$ is the smallest root of
$u^{2}$. This condition on the period of $\phi$ is due to the
requirement of smoothness of the metric at the bolt.

 In the Lorentzian case where $\phi $ is unwrapped to $t$, there is no
such restriction and the full 1-parameter family of Schwarzschild
metrics $g_{\mathrm{Schw}}(m)$ has the {\it  same}  boundary
metric, i.e.
$$- dt^{2} + g_{S^{2}(1)}.$$
The same feature essentially holds for general \gstatic\ metrics on
$M$ in \eq{e2.9}. Namely, choose a Riemannian static boundary
metric in Im $\Pi_{S}$,
$$\gamma  = \gamma^{R} = \alpha^{2}d\phi^{2} + g_{S^{2}}. $$
Suppose a neighborhood of $[\gamma]$ in ${\cal C}_{S}$ is also in
Im $\Pi_{S}$, cf. the remarks above on the structure of the variety
$\Pi_{S}({\cal E}_{S}) \subset {\cal C}_{S}$. In this case, one can vary
$\alpha $ by $\alpha \rightarrow (1+s)\alpha$, for
$s\in (-\varepsilon , \varepsilon )$, while remaining in Im $\Pi_{S}$.
This gives a curve of distinct boundary metrics $\gamma_{s}^{R}$, or more
precisely, a non-trivial curve in the space of conformal classes
${\cal C}_{S}$. This curve gives rise to a curve of
isometrically distinct filling Einstein metrics $g_{s}^{R}$ in the Riemannian
case. Hence, one also has a curve $g_{s}^{L}$ of isometrically distinct
Lorentzian-Einstein metrics.

  However, the corresponding curve of Lorentzian boundary metrics
$\gamma_{s}^{L}$ are all isometric, in that there is a curve of
diffeomorphisms $\psi_{s}$ of $\partial M$, ($\psi_{s} \neq id$),
such that $(\psi_{s}^{L})^{*}\gamma_{0}^{L} = \gamma_{s}^{L}$.
Namely,
$$\gamma_{s} = -\alpha^{2}(1+s)^{2}dt^{2} + g_{S^{2}} =
\psi_{s}^{*}(\gamma_{0}), $$ where $\psi_{s}(t,q) = ((1+s)t,q)$,
 $q\in S^{2}.$

 Thus,  if one identifies such boundary metrics differing only
by a dilation of the $t$ factor, one then has local 1-parameter
 families of
distinct AdS \gstatic\ vacuum Lorentz metrics with the same boundary
metric. Presumably, the 1-parameter comes from an appropriate
notion of mass.

\medskip

 Regarding uniqueness, we have the following analogue of the black hole
uniqueness theorem when $\Lambda = 0$, or analogue of the rigidity of
anti-de Sitter space in Proposition 3.2.

\begin{proposition} \label{p 2.3.}
  The AdS Schwarzschild metrics $g_{\mathrm{Schw}}(m)$ are the unique
  \gstatic\
 vacuum metrics on ${\Bbb R}\times ({\Bbb R}^{+}\times S^{2}),$
smooth up to the horizon, which admit a $C^{2}$ (or $L^{2,p}, p >
$ 4) conformal completion, with a conformally compactifiable smooth acausal
hypersurface orthogonal to the static Killing field, with boundary metric
$\gamma_{0} = - dt^{2} + g_{+1},$ where $g_{+1}$ is the round metric on
$S^{2}.$
\end{proposition}

\noindent \proof The proof is the same as that of Proposition 3.2.
Thus, it suffices to prove this in the Riemannian setting, and
this has already been done in \cite[\S 7]{mand2}, using the
fact that any AH Einstein filling metric has an isometric
$S^{1}\times SU(2)$ action. {\endproof}

Propositions~\ref{p 2.2.} and \ref{p 2.3.} have the following
important corollary:

\begin{theorem}\label{tau}
 The standard AdS metric and the domain of outer communication of the
 AdS Schwarzschild metrics $g_{\mathrm{Schw}}(m)$, $m>0$, are the unique
 static
globally hyperbolic (in the sense of manifolds with boundary)
vacuum metrics which admit $C^{2}$ (or $L^{2,p}, p
> $ 4) conformal completions,
with conformally compactifiable acausal spacelike surfaces (with
perhaps an interior boundary on which $u$ vanishes), without
degenerate event horizons, and with boundary metric (at infinity)
 $\gamma_{0}
= - dt^{2} + g_{+1},$ where $g_{+1}$ is the round metric on
$S^{2}.$
\end{theorem}

\proof Global hyperbolicity implies that $\lormet$ has the global
representation \eq{e2.6}, so that the Riemannian counterpart of
$(\mcM,\lormet)$ satisfies all the hypotheses of Section~\ref{S1}.
(The non-degeneracy of the event horizons ensures compactness of
the associated Riemannian manifold,
compare~\cite[Section~3]{Chstatic}.) The large isometry
group forces the topology of the filling space-time to be that
considered in Propositions~\ref{p 2.2.} or \ref{p 2.3.}, and the
result immediately follows. \qed

 As
usual the above claim concerns only the domain of outer
communication. No natural reasonable conditions are known which
would guarantee some form of uniqueness beyond the event horizon.
We note that there are at least two ways of adding a boundary
$\{u=0\}$ to the set $\{u>0\}$ on the Cauchy surface: either by
gluing in a sphere (which becomes a boundary for the new
manifold), or by gluing in a $\R{\Bbb P}^2$ (which becomes an
interior surface for the new manifold).

\subsection{$V = T^{2}.$}

 Then $\partial \mcM = {\Bbb R}\times T^{2},$ or $\partial M =S^{1}\times T^{2}=T^3$ in
the Riemannian setting. First consider the Riemannian setting. As
filling manifold, we choose
\begin{equation} \label{e2.13}
M = D^{2}\times T^{2},
\end{equation}
where $D^{2}$ is a disc filling  {\it some} $S^{1} \subset T^{3} =
\partial M$.  This manifold carries the ``generalised Riemannian AdS
$T^{2}$ black hole"
%\ptc{I have put quotes around the name, the
%"black hole" sounds kind of funny to me in this context}
metrics
\begin{equation} \label{e2.14}
g_{T^{2}}(m) =  [u^{-2}dr^{2} + (u^{2}ds^{2} +
r^{2}\gamma_{0})]/\Gamma ,
\end{equation}
where $u^{2} = (r^{2} -  \frac{2m}{r})$, $m > 0$, $\gamma_{0}$ is
the flat metric on ${\Bbb R}^{2},$ and $\Gamma $ is any flat
lattice in ${\Bbb R}^{3} = {\Bbb R}\times {\Bbb R}^{2}$. Here
$\Gamma$ is arbitrary, subject to the constraint that the $S^{1}
\subset {\Bbb R}^{3}/\Gamma$ bounding the disc $D^{2}$ in
\eq{e2.13} has $\gamma_0$--length depending on $m$, to give a
smooth metric where $u = 0$; this is the same issue as the
constraint on $\beta$ following \eq{e2.12}. Alternately,
prescribing the $\gamma_0$--length determines $m$. The action of
$\Gamma$ on ${\Bbb R}^{3}$ has many distinct extensions to an
action of $\Gamma$ on the universal cover $\widetilde M =
D^{2}\times {\Bbb R}^{2}$, cf. \cite[\S 4.3 and 4.4]{mand1} for
further details.

  A simple computation shows that the metrics \eq{e2.14} all have
non-positive sectional curvature and hence they are regular points
of $\Pi$, cf. \cite{Lee:fredholm,Biquard}. In fact, the metrics
$g_{T^{2}}$ are all locally isometric.  Thus, as before,
$\Pi({\cal E})$ is a variety of codimension 0 in ${\cal C}$.

However, in general, metrics of the form \eq{e2.14} are not
\gstatic\ w.r.t. any $S^{1}$ action - they are only locally
static. Moreover, they are locally static w.r.t. many $S^{1}$
actions. Because the proof uses global arguments, we do not know
if Theorem~\ref{t 1.2} holds for locally static $S^{1}$ actions on
$\partial M$, (although we expect this to be the case).

%  Thus, in the following, we consider only those lattices $\Gamma =
%\Gamma_{S}$ which preserve {\it some} orthogonal product decomposition
%${\Bbb R}^{3} = {\Bbb R}\times {\Bbb R}^{2}$, i.e.
%$\Gamma_{S} = {\Bbb Z}\times \Gamma_{o}$, $\Gamma_{o}$ a lattice on
%${\Bbb R}^{2}$. The corresponding (non-generalised) $T^{2}$ black hole
%metrics \eq{e2.14} are \gstatic\ w.r.t. rotation in the associated
%$S^{1} = {\Bbb R}/{\Bbb Z}$ factor.
%
%  Hence, as before, the full component ${\cal E}_{S}^{c}$ of ${\cal E}_{S}$
%containing any such $g_{T^2}(m)$ consists of \gstatic\ AH Einstein
%metrics on $M$.

  In this situation, ${\cal C}^{o} = \emptyset $ -- there are no
non-flat metrics of non-negative scalar curvature on $\partial M =
T^{3}$, and the degree theory breaks down. In fact the boundary
map $\Pi$ is {\it not} proper on flat boundary metrics
$\gamma_{0}$ on $T^{3} = S^{1}\times T^{2}$; there are infinitely
many distinct AH Einstein metrics $g_{i} = g_{i}(\gamma_{0})$, all
of the form \eq{e2.14} with a given conformal infinity
$(S^{1}\times T^{2}, \gamma_{0}),$ for any fixed flat metric
$\gamma_{0}$. The infinite sequence comes from the infinite number
of ways $S^{1} \subset \partial M$ may bound a disc $D^{2}$ in
\eq{e2.13}, or equivalently the infinite number of extensions of a
given $\Gamma$ action to an action on $\widetilde M$. We point out
that these metrics $g_{i}$ lie in distinct components ${\cal
E}_{S}^{c}$ of the full moduli space ${\cal E}_{S}$ in \eq{e2.2}.
Thus, ${\cal E}_{S}$ in fact has infinitely many distinct
components ${\cal E}_{S}^{c}$, each consisting of static AH
Einstein metrics on $M$, cf. again \cite[\S 4.4]{mand1}, for
further details.

  Because the degree theory breaks down, there is currently no general
existence theorem for AH Einstein metrics on $M$ with prescribed
conformal infinity, as in Corollary~\ref{c 2.1.}. Nevertheless, as
mentioned above, each component ${\cal E}_{S}^{c}$ of ${\cal
E}_{S}$ containing a $T^{2}$ black hole metric is a large,
infinite dimensional space, with $\Pi({\cal E}_{S}^{c})$
containing open sets of ${\cal C}_{S}$.

  We turn to the Lorentzian situation below, but first prove the following
uniqueness theorem for Riemannian AH Einstein metrics with a flat
boundary metric on $T^{3}$.
%This result is of course an analogue
%of Propositions~\ref{p 2.2.} and \ref{p 2.3.} in this setting.

\begin{proposition} \label{p 2.4.}Let $\gamma_{0}$ be a flat metric on
$\partial M = T^{3} = S^{1}\times S^{1}\times S^{1}$. Then any AH
Einstein metric with conformal boundary class $[\gamma_{0}]$ is of
the form~\eq{e2.14}. Further, for every $\gamma_0$ there exists
precisely a countable infinity of distinct filling AH Einstein
metrics~\eq{e2.14}, each member of this family being uniquely
determined by specifying which $S^{1} \subset T^{3}$, (not
necessarily a product factor), bounds the disc $D^{2}$ in
\eq{e2.13}.
\end{proposition}

 \noindent \proof Any AH Einstein metric
$g$ on a manifold $N$ with conformal infinity a flat metric on
$T^{3}$ has an effective isometric $T^{3}$ action, by
Proposition~\ref{p 1.1}. In particular, the principal orbits of
this action are all flat tori, and so the manifold $N$ must be of
the form $M$ in \eq{e2.13}.  On the universal cover the metric
takes thus the form
$$ g_{rr}(r)dr^2 + g_{ij}(r)(dx^i + U^i(r)dr)(dx^j+U^j(r)dr)\;,$$
with $r$ in an interval $I\subset \R$. We wish to show that
without loss of generality the metric can be taken to be
diagonal\footnote{The argument given here is a straightforward
rephrasing of the usual argument that Bianchi I metrics are
diagonal.}. Choose $r_0\in I$, a change of coordinates $x^i\to
y^i=x^i+\int _{r_0}^rU^i(s)ds$ sets the $U^i$'s to zero. A
reparameterization of $I$ leads to $g_{rr}=1$. By a linear change
of coordinates we can achieve $g_{ij}(r_0)=\delta_i^j.$ A further
($r$--independent) rotation of the coordinates leads to
$\partial_r g_{ij}{(r_0)}=\mbox{\rm diag}(a_0^2, b_0^2,c_0^2)$.
Now, the Einstein equation for $g$ implies a second order system
of ODE's for $g_{ij}$ (for the reader familiar with general
relativity, these are the usual ADM evolution equations with zero
shift and with the square of the lapse function replaced by $-1$),
giving a unique solution for every set of strictly positive
numbers $a_0,b_0,c_0$; this set will be referred to as ``set of
equations GE". (Here we do not normalize the scalar curvature,
demanding only  strict negativity, so that there is no ``scalar
constraint equation".) On the other hand, the equations which are
obtained by assuming that the metric is diagonal (see
e.g.~\cite{Linet}) show that there also exists a solution with
$g_{ij}(r)$ --- diagonal for all $r$. Uniqueness of solutions of
the ``set of equations GE" implies that in the above coordinate
system $g_{ij}(r)$ is indeed diagonal.
  It
follows that there exists a coordinate system in which the metric
$g$ is of the form
$$dr^{2} + a^{2}(\theta^{1})^{2} + b^{2}(\theta^{2})^{2} +
c^{2}(\theta^{3})^{2}\;,$$ where $a$, $b$, $c$ are functions of
$r$ only, while the $\theta^i$'s form a constant coefficients
basis of the space of one-forms anihilating $\partial_r$. It is
now a straightforward computation to see that the only such
metrics which give a smooth Einstein conformally compactifiable
metric on $M$ are of the form \eq{e2.14}
(compare~\cite{Linet,Kasner}). This justifies uniqueness.

As in the discussion prior to the proposition, there is a
countable infinity of possible choices of which simple closed
geodesic $\Gamma\subset T^3 $ will bound the disc $D^2$ occurring
in \eq{e2.13}, and each such choice will lead to a different
filling metric, uniquely determined by $\Gamma$, as
follows:\footnote{An alternative way of understanding the
non-uniqueness is as follows: any flat metric $\gamma$ on $T^3$
can be represented by a constant coefficient matrix
$\gamma=\gamma_{ij}d\varphi^i d\varphi^j$. The representation is
unique up to a linear relabeling of the $\varphi^i$'s, $\varphi^i
\to L^i{}_j\varphi^j$, with $L^i{}_j\in SL(3,\Z)$. If we decree
that the $D^2$ factor in \eq{e2.13} fills the first $S^1$ factor
of $T^3$, then  each choice of $L^i{}_j$ will lead to a different
filling metric.}
 let $\gamma_0$ be an arbitrary flat metric on
$T^3$, and let $\Gamma\approx S^1\subset T^3$ be the simple closed
geodesic which will be filled by a $D^2$. Let $\varphi\in
[0,2\pi]$ be a $2\pi$ periodic affine parameter on $\Gamma$, and
let $(\varphi_1, \varphi_2,\varphi_3)\in[0,2\pi]^3$ be a
coordinate system on $T^3$ such that $\varphi_1|_\Gamma\equiv
\varphi$, $\varphi_2|_\Gamma=\varphi_3|_\Gamma=0$, and  such that
the one-forms $d\varphi_i$ are $\gamma_0$--covariantly constant.
Let $\theta^i$ be a $\gamma_0$--orthonormal co-frame on $T^3$ with
$\theta^2(\partial_{\varphi_1})=\theta^3(\partial_{\varphi_1})=0$.
Rescaling $\gamma_0$ by a constant factor if necessary we can
assume that the $\gamma_0$--length of $\Gamma$ equals $2\pi$, thus
$\theta^1(\partial_1)=1$. Parameterising $D^2$ by polar
coordinates $(\rho,\varphi_1)\in[0,1)\times[0,2\pi]$ we extend the
$\theta^i$'s to $D^2\times T^2$ so that
$\mcL_{\partial_\rho}\theta^i = \theta^i(\partial_\rho)=0.$ Let
$r:[0,1)\to[1/2,\infty)$ be any diffeomorphism of $[0,1)$ with
$[1/2,\infty)$, then the desired smooth filling metric is
$$g=u^{-2}dr^2 + u^2 (\theta^1)^2 +r
^2((\theta^2)^2+(\theta^3)^2)\;,$$ with
$$u^2(r)=r^2 - 2m/r\;, \qquad m= 2^{-4}\;.$$
{\endproof}

 Now we turn to the Lorentzian setting, still assuming that $\partial M =
S^1 \times T^{2}$, so that $\partial \mcM = {\Bbb R} \times
T^{2}$.  Suppose first that the associated Lorentzian space-time
$\mcM$ takes the form
\begin{equation} \label{e2.15}
\mcM = {\Bbb R}\times (D^{2}\times S^{1})\;,
\end{equation}
with $M =S^1\times (D^{2}\times S^{1})$. Here the static action on
$\partial M$ comes from a free \gstatic\ $S^{1}$ action on the
$T^{2}$ factor in \eq{e2.13}. Since the $S^{1}$ orbits do not
bound a disc in $M$, it follows that the $S^{1}$ action is
\sgstatic. Hence, the corresponding Lorentz metric on $\mcM$ as in
\eq{e2.15} is a \sgstatic\ vacuum AdS metric and so has the form
\begin{equation} \label{e2.16}
g = - u^{2}dt^{2} + g_{\Sigma},
\end{equation}
where $\Sigma  = D^{2}\times S^{1}$ and $u > 0$ on $\Sigma$.

  The same remarks on the existence of a large space of such vacuum AdS
solutions as in the Riemannian case apply here also.

 The main examples in this situation are the Horowitz-Myers AdS soliton
metrics, \cite{HorowitzMyers}, in dimension 4:
\begin{equation} \label{e2.17}
g_{T^{2}}(m) =  - r^{2}dt^{2} + u^{-2}dr^{2} + (u^{2}ds_{1}^{2}+
r^{2}ds_{2}^{2})/\Gamma_{0},
\end{equation}
where $u^{2} = (r^{2} - \frac{2m}{r}),$ and $\Gamma_{0}$ is any
lattice acting on ${\Bbb R}^{2} = {\Bbb R}^{2}(s_{1},s_{2})$.

 The Riemannian--Lorentzian correspondence and Proposition 3.5 then imply
immediately the following uniqueness result for the AdS soliton
metrics. This result generalises a recent result of Galloway,
Surya and Woolgar~\cite{Galloway:2001uv,Galloway:2002ai}, in
dimension four. We note that the $SL(3,\Z)$ discrete degrees of
freedom of Proposition~\ref{p 2.4.} are reduced now to $SL(2,\Z)$
discrete degrees of freedom:

\begin{proposition} \label{p 2.5.} All
   conformally compactifiable static vacuum AdS
metrics on $\R\times D^2\times S^1$ with acausal conformally
compactifiable hypersurfaces orthogonal to the static Killing
vector and with (conformal) boundary metric the flat product $-
dt^{2} + \gamma_{0}$ on ${\Bbb R}\times T^2$ are of  the form
\eq{e2.17}. Further, for any such $\gamma_0$ there is precisely a
countable family of filling Einstein metrics \eq{e2.17}, each
member of this family being determined by specification of the
$S^{1} \subset T^{2}$ (not necessarily a factor of the product
$T^2=S^1\times S^1$) bounding the disc in \eq{e2.15}.
\end{proposition}
{\endproof}

% Again, $\Gamma_{0}$ depends on $m$, as in the Riemannian case. Hence,
%the boundary metrics $\gamma$ on ${\Bbb R}\times T^{2}$ vary as $m$ varies.

 It seems of interest to make
some comments about cylindrically symmetric, or axially symmetric,
\sgstatic\ metrics. Let $\lormet$ by any of the vacuum Lorentzian
metrics on $\R\times D^2\times S^1$ mentioned after \eq{e2.16}.
Unwinding the last   $S^1$--factor and calling $z$ the associated
coordinate along $\R$ one obtains a $z$-periodic solution on
$\R\times D^2\times \R=\R\times \R^3$. If the boundary metric is
invariant under rotations of the $S^1=\partial D^2$ factor, then
so is the filling metric, and we have thus obtained a
$z$--periodic axi-symmetric solution. If, instead, the boundary
metric is invariant under rotations along the last  $S^1$--factor,
the resulting filling metric will be invariant under
$z$--translations. It is a corollary of Proposition~\ref{p 2.5.}
that all appropriately regular cylindrically symmetric  \sgstatic\
solutions in the class considered here belong to the
Horowitz-Myers family; the latter correspond to the cylindrically
symmetric strings of Linet~\cite{Linet} with the constant $K$ in
\cite{Linet} equal to one.

 Next, we turn to the case where there is a horizon. In this situation,
the $S^{1}$ action on $\partial M$ is taken to be rotation in the
$S^{1}$ factor bounding $D^{2}$. Hence, the Lorentzian manifold is
of the form
\begin{equation} \label{e2.18}
\mcM = {\Bbb R}\times ({\Bbb R}^{+}\times T^{2}).
\end{equation}
The metric is of the form \eq{e2.16}, but with a non-empty
${\Bbb R}\times T^{2}$ horizon $\{u = 0\}$. Again the same remarks on the
existence of a large infinite dimensional family of such $T^{2}$ AdS black
hole metrics apply as before.

 Similarly, the Riemannian--Lorentzian correspondence and Proposition 3.5 give
immediately the following uniqueness theorem. Note that the
discrete degrees of freedom of Proposition~\ref{p 2.4.} have
disappeared, since the $S^1$ bounding a disc has been chosen to
correspond to the $\R$ factor:

\begin{proposition} \label{p 2.6.old}
  The AdS $T^{2}$ black hole metrics
\begin{equation} \label{e2.19}
g_{T^{2}}(m) =  - u^{2}dt^{2} + u^{-2}dr^{2} +
(r^{2}\gamma_{0})/\Gamma_{0},
\end{equation}
where $u^{2} = (r^{2} - \frac{2m}{r}), \gamma_{0}$ is the flat
metric on ${\Bbb R}^{2},$ and $\Gamma_{0}$ is any lattice acting
on ${\Bbb R}^{2} = {\Bbb R}\times {\Bbb R}$, are the unique
\gstatic\ vacuum conformally compactifiable AdS metrics with a
smooth acausal conformally compactifiable hypersurface orthogonal
to the static Killing field (with a boundary on which $u=0$) and
with boundary metric the flat metric $- dt^{2} + {\Bbb
R}^{2}/\Gamma_{0}.$
\end{proposition}
{\endproof}

 As in the $S^{2}$ black hole case, as $m$ varies, the boundary metrics are
fixed, for a given lattice.

Similarly to Theorem~\ref{tau}, we obtain:

\begin{theorem}\label{tau2}
 The Horowitz-Myers metrics and the domains of outer communications of the toroidal Kottler metrics
  are the only static globally hyperbolic (in the sense of space-times with
boundary) vacuum metrics which admit  $C^{2}$ (or $L^{2,p}$, $p >
4$) conformal completions, with conformally compactifiable Cauchy
surfaces (with perhaps an interior boundary on which $u$
vanishes), without degenerate event horizons, and with boundary
metric (at infinity) $\gamma_{0} = - dt^{2} + g_{0},$ where
$g_{0}$ is a flat metric on $T^{2}$. In the strictly static case
there is precisely one distinct filling Horowitz-Myers metric for
each choice  of an $S^1\subset T^2$ (not necessarily a product
factor) which bounds a disc $D^2$ in the filling space-time.
\end{theorem}

A possible interpretation of Theorem~\ref{tau2} is  that metrics
as above are classified (up to discrete degrees of freedom) by
their global (Hamiltonian) mass $M$: negative $M$ leads to the
Horowitz-Myers strictly static solutions, positive $M$ leads to
the Kottler black hole solutions, solutions with zero $M$ do not
exist. Note that $M$ is proportional to the parameter $m$
appearing in \eq{e2.19} in the black hole case, and proportional
to minus the parameter $m$ appearing in \eq{e2.17} in the strictly
static case. In the Horowitz-Myers case this point of view is
somewhat misleading, as the mass itself is actually determined by
 $[\gamma_0]$ and the choice of the filling disc: thus, to every
 boundary class $[\gamma_0]$ one can uniquely associate a discrete sequence
of masses.\footnote{The construction of the proof of
Proposition~\ref{p 2.4.} gives $m=2^{-4}$ for all metrics, in the
normalisation of the boundary metric in which the bounding $S^1$
has prescribed length. However, when defining mass with $V=T^2$
one needs to normalise the $\gamma_0$--area of $T^2$; this will
lead to distinct masses of the filling metrics once some such
normalisation has been chosen.}

%A somewhat larger class of locally static
%space-times is obtained by considering metrics of the form
%\begin{equation} \label{e2.19n}
%g_{T^{2}}(m) =  - u^{2}(dt+\alpha)^2  + u^{-2}dr^{2} +
%(r^{2}\gamma_{0})/\Gamma_{0},
%\end{equation}
%where $\alpha$ is  one-form on $T^2$ which has constant
%coefficients in the natural coordinates there, and $u=u(m)$ is as
%above. This is the filling metric which one obtains for a general
%flat Lorentzian boundary metric on $\R\times T^2$. Our uniqueness
%Theorem~\ref{tau2} generalises immediately to such metrics.

\subsection{$V = \Sigma_{g}$, $g \geq  2$.}

 Then $\partial M = {\Bbb R}\times \Sigma_{g}$, $S^{1}\times \Sigma_{g}$
in the Riemannian setting. Again, there exist no metrics of
non-negative scalar curvature on $\partial M$, and so there is no degree
theory.

  There are no known \sgstatic\ AH Einstein metrics on
${\Bbb R}\times H_{g}$, where $H_{g}$ is a 3-manifold with
$\partial H_{g} = \Sigma_{g}$; (for instance $H_{g}$ a
handlebody). It has been conjectured  that in fact no such
\sgstatic\ solutions exist.

  On the other hand, suppose
\begin{equation} \label{e2.20}
M = D^{2}\times \Sigma_{g}.
\end{equation}
On such an $M$, one has the Riemannian genus $g$ AdS black hole metrics
\begin{equation} \label{e2.21}
g = u^{2}d\phi^{2} + u^{-2}dr^{2} + r^{2}g_{- 1},
\end{equation}
where $u^{2} = - 1+r^{2} -  \frac{2m}{r},$ and $g_{-1}$ is a
hyperbolic metric on $\Sigma_{g}$. (As before, the period of
$\phi$ depends on $m$). These metrics are \gstatic, and so the
full component ${\cal E}_{S}$ containing such metrics consists of
\gstatic\ AH Einstein metrics on $M$. Each of these metrics has
negative sectional curvature and so is a regular point of $\Pi$,
as in the $T^{2}$ case. As before then, Im $\Pi_{S}$ is a variety
of codimension 0 in ${\cal C}_{S}$. Thus, again one has the
existence of large, infinite dimensional space of such AH Einstein
metrics, and a correspondingly large space of Lorentzian \gstatic\
AdS vacuum genus $g$ black hole metrics. These metrics are of the
form \eq{e2.16} with $\Sigma = \Sigma_{g}$ and with horizon $\{u =
0\} = {\Bbb R}\times \Sigma_{g}$.

  A uniqueness theorem as in Propositions~\ref{p 2.2.}, \ref{p 2.3.}, etc. for such metrics
with constant curvature boundary metric on $\Sigma_{g}$ remains
currently an open problem however. This is because the symmetries
of $(\Sigma_{g}, g_{-1})$ are only local symmetries, and the
analogue of Proposition~\ref{p 1.1} in this context, (i.e. for
local symmetries), remains open.

\section{Properties of the Lorentzian solutions}
\label{Sglobal}

Consider any \sgstatic\ Lorentzian metric as constructed in the
previous section, thus we have $\mcM=\R\times \Sigma$ with the
vacuum metric \bel{e0} \lormet = -u^2dt^{2}+\threeg\;,\ee as in
\eq{e2.6}. We wish to justify the properties of $(\mcM,\lormet)$
listed in the Introduction. Now, point~\ref{p1} follows by
construction. Point~\ref{p5} follows from what has been said in
the proof of Theorem~\ref{t 1.2}. Point~\ref{p2} is a
straightforward corollary of point~\ref{p5}.   The geodesic
completeness of the static metrics so obtained is
 a special case of the following result:

\begin{Theorem}\label{Tgc} Let $\bS=
\partial \Sigma \cup\Sigma$ be a compact  manifold with boundary.
Consider a space-time $(\mcM=\R\times \Sigma,\lormet)$ such that
\bel{gc1} \lormet = -V^2dt^2+\threeg \;, \quad \mLX V= \mLX
\threeg  = \threeg (X,\cdot)=0\;,\ee $X=\partial_t$,  with $V$
strictly bounded away from zero. Suppose that $V^{-2} \threeg $
can be extended by continuity to a differentiable Riemannian
metric $\bg$ on  $\bS$, such that $\pS$ is totally geodesic in
$(\Sigma, \bg)$. We also assume that $V^{-1}$ can be extended by
continuity to a differentiable function $\Omega$ on $\bS$, with
$\Omega$ vanishing precisely on $\pS$ and  $d\Omega$ nowhere
vanishing on $\pS$. Then $(\mcM,\lormet)$ is
 geodesically complete.
\end{Theorem}

\begin{Remark} The hypotheses of Theorem~\ref{Tgc} are satisfied by all
the metrics obtained by ``Wick rotation" of  metrics considered in
the proof of Theorem~\ref{t 1.2}.
\end{Remark}

 \proof Along any maximally extended affinely parameterised geodesic
$\Gamma(s)=(t(s),\gamma(s))$ we have
$$\frac d  {ds}\left(\lormet(X,\dot\Gamma)\right)=0\;,\qquad
\lormet(\dot\Gamma,\dot\Gamma)=:\eta\in\{0,\pm 1\}\;.$$ Let
$-\mcE$ denote $\lormet(X,\dot\Gamma)$, then $d\mcE/ds=0$ and
\beal{gc3} \frac {dt}{ds} &=& \frac {\mcE}{V^2} \;,
\\ \threeg (\dot \gamma,
\dot \gamma) &=& \eta + \frac {\mcE^2}{V^2}\;. \eeal{gc5} Since
$V$ is strictly bounded away from zero, \eq{gc3} implies that $t$
cannot blow up along $\Gamma$ in finite affine time $s$.

For spacelike geodesics \eq{gc5} implies that $\threeg(\dot
\gamma,\dot \gamma)$ is uniformly bounded, and completeness of
$\Gamma$ follows from the geodesic completeness of
$(\Sigma,\threeg)$.

 For timelike geodesics we have $\eta=-1$ and
since $\threeg (\dot \gamma,\dot\gamma)\ge 0$, \eq{gc5}  shows
that
$$V^2\le \mcE^2$$ along $\Gamma$. It follows that $\Gamma$ stays
away from the conformal boundary $\R\times\pS$, and timelike
geodesic completeness easily follows.

Null geodesics of $\lormet$ are also null geodesics of the
\emph{optical metric}
$$-dt^2+V^{-2}\threeg \;.$$
The proof that any null geodesic which stays away from
$\partial\Sigma$ is complete proceeds as in the timelike case. It
remains to consider  a null geodesic $\Gamma$ which approaches
$\partial \Sigma$. Let $\hat\Gamma(r)=\Gamma(s(r))$ be a reparameterization
of $\Gamma$ so that $r$ is an affine parameter with respect to the optical
metric. Then $r$ can be chosen so that $\hat\Gamma(r)=(r,\hat\gamma(r))$,
where $\hat \gamma$ is a unit-speed geodesic of the metric $\bg$. By hypothesis
$\hat\gamma$ meets $\pS$ in finite time. The totally geodesic
character of $\pS$ implies that $\hat\gamma$ has to meet $\pS$
transversally, and the completeness of $\Gamma$ follows from the
usual formula for the change of affine parameter of null geodesics
under conformal rescalings of the metric. \qed

Returning to the justification of our claims from the
introduction, we note  the well known fact that ON-tetrad
components of any finite number of covariant derivatives of the
Riemann tensor are uniformly bounded on $M$ for all Riemannian
conformally compactifiable metrics. In the obvious tetrad
associated with the representation \eq{e2.3} of $\riemg$, the
algebraic invariants of the associated Lorentzian metric $\lormet$
will be polynomials with constant coefficients of those tetrad
components, and point~\ref{p4} immediately follows. Global
hyperbolicity in point~\ref{p5a} is established in the course of
the proof of Theorem~4.1 of~\cite{ChruscielSimon}.

It remains to consider the question of existence of other, perhaps
only locally defined, Killing vector fields. Consider a space-time
$(\mcM,\lormet)$ as in Theorem~\ref{Tgc} and suppose that there
exists an open subset of $\mcM$ with a Killing vector field $Y$
defined there, such that $Y$ is not proportional to $X$. Now, it
is well known that $\Sigma$ admits an analytic
atlas with respect to which $\lormet$ is analytic. This allows us to make
use of a theorem of Nomizu~\cite{Nomizu}, and conclude that $Y$ can be
extended to a globally defined Killing vector field on the universal cover
$\R\times\hat\Sigma$ of $\R\times\Sigma$. In particular,  for any
point $p\in \pS$ there exists a $\pS$--neighborhood $\mcU$ of $p$
such that the algebra of Killing vector fields defined  on
$\R\times(\mcU\setminus\pS)$ is at least two-dimensional. Now,
Killing vector fields are conformal vector fields of the
conformally rescaled metric. Further, conformal Killing vector
fields satisfy an overdetermined system of ODE's of third order,
which easily implies that they extend to the conformal boundary as
conformal Killing vectors there. This can happen only for very
special  $\alpha$'s and $g_V$'s in \eq{e2.1}, which justifies our
claim \ref{p7} in the introduction. Point~\ref{p6} is an obvious
corollary of point~\ref{p7}.

Identical arguments apply to those static solutions which have
zeros of $u$, leading to black hole space-times.

\medskip

\noindent{\sc Acknowledgements:} PTC acknowledges useful
discussions with Witold Kondracki and  Abdelghani Zeghib.

\def\cprime{$'$}
\providecommand{\bysame}{\leavevmode\hbox
to3em{\hrulefill}\thinspace}
\providecommand{\MR}{\relax\ifhmode\unskip\space\fi MR }
\providecommand{\MRhref}[2]{%
  \href{http://www.ams.org/mathscinet-getitem?mr=#1}{#2}
} \providecommand{\href}[2]{#2}


\begin{thebibliography}{10}

\bibitem{manderson:stationary}
M.T. Anderson, \emph{On stationary vacuum solutions to the
{E}instein
  equations}, Annales H.~ Poincar\'e \textbf{1} (2000), 977--994,
  gr-qc/0001091.

\bibitem{manderson:static}
\bysame, \emph{On the structure of solutions to the static vacuum
{E}instein
  equations}, Annales H.~ Poincar\'e \textbf{1} (2000), 995--1042,
  gr-qc/0001018.

\bibitem{mand1}
\bysame, \emph{Boundary regularity, uniqueness and non-uniqueness
for {AH
  Einstein} metrics on $4$-manifolds}, Adv.\ in Math. (2001), in press,
  math.DG/0104171.

\bibitem{mand2}
\bysame, \emph{{Einstein} metrics with prescribed conformal
infinity on
  $4$-manifolds},  (2001), math.DG/0105243.

\bibitem{AndDahl}
L.~Andersson and M.~Dahl, \emph{Scalar curvature rigidity for
asymptotically
  locally hyperbolic manifolds}, Annals of Global Anal.\ and Geom. \textbf{16}
  (1998), 1--27, dg-ga/9707017.

\bibitem{VanderBij:2001ia}
J.J.~van~der Bij and E.~Radu, \emph{New hairy black holes with
negative
  cosmological constant}, Phys.\ Lett. \textbf{B536} (2002), 107--113,
  gr-qc/0107065.

\bibitem{Biquard}
O.~Biquard, \emph{{M\'etriques d'Einstein asymptotiquement
sym\'etriques
  (Asymptotically symmetric Einstein metrics)}}, {Ast\'erisque 265, Paris:
  Soci\'et\'e Math\'ematique de France, 109 pp.}, 2000.

\bibitem{Bjoraker:2000qd}
J.~Bjoraker and Y.~Hosotani, \emph{Monopoles, dyons and black
holes in the
  four-dimensional {Einstein-Yang-Mills} theory}, Phys. Rev. \textbf{D62}
  (2000), 043513.

\bibitem{BGH}
W.~Boucher, G.W. Gibbons, and G.T. Horowitz, \emph{Uniqueness
theorem for
  anti--de {S}itter spacetime}, Phys.\ Rev.\ D \textbf{30} (1984), 2447--2451.

\bibitem{ChAscona}
P.T. Chru\'sciel, \emph{Uniqueness of black holes revisited},
Helv. Phys. Acta
  \textbf{69} (1996), 529--552, Proceedings of Journ\'es Relativistes 1996,
  Ascona, May 1996, N. Straumann,Ph. Jetzer and G. Lavrelashvili (Eds.),
  gr-qc/9610010.

\bibitem{Chstatic}
\bysame, \emph{The classification of static vacuum space--times
containing an
  asymptotically flat spacelike hypersurface with compact interior}, Class.
  Quantum Grav. \textbf{16} (1999), 661--687, gr-qc/9809088.

\bibitem{ChHerzlich}
P.T. Chru\'sciel and M.~Herzlich, \emph{The mass of asymptotically
hyperbolic
  {R}iemannian manifolds},  (2001), dg-ga/0110035.

\bibitem{ChruscielSimon}
P.T. Chru\'sciel and W.~Simon, \emph{Towards the classification of
static
  vacuum spacetimes with negative cosmological constant}, Jour.\ Math.\ Phys.
  \textbf{42} (2001), 1779--1817, gr-qc/0004032.

\bibitem{Fintushel2}
R.~Fintushel, \emph{{Circle actions on simply connected
4-manifolds.}}, Trans.
  Am. Math. Soc. \textbf{230} (1977), 147--171.

\bibitem{Fintushel1}
\bysame, \emph{{Classification of circle actions on
4-manifolds.}}, Trans. Am.
  Math. Soc. \textbf{242} (1978), 377--390.

\bibitem{Galloway:2002ai}
G.J. Galloway, S.~Surya, and E.~Woolgar, \emph{On the geometry and
mass of
  static, asymptotically {AdS spacetimes, and the uniqueness of the AdS}
  soliton},  (2002), hep-th/0204081.

\bibitem{Galloway:2001uv}
\bysame, \emph{A uniqueness theorem for the {AdS} soliton}, Phys.\
Rev.\ Lett.
  \textbf{88} (2002), 101102, hep-th/0108170.

\bibitem{GibbonsHawkingInstantons}
G.~W. Gibbons and S.~W. Hawking, \emph{Classification of
gravitational
  instanton symmetries}, Commun. Math. Phys. \textbf{66} (1979), 291--310.

\bibitem{GL}
C.R. Graham and J.M. Lee, \emph{Einstein metrics with prescribed
conformal
  infinity on the ball}, Adv. Math.\ \textbf{87} (1991), 186--225.

\bibitem{Hawking:1983dh}
S.~W. Hawking and Don~N. Page, \emph{Thermodynamics of black holes
in anti-de
  sitter space}, Commun. Math. Phys. \textbf{87} (1983), 577.

\bibitem{Heusler:book}
M.~Heusler, \emph{Black hole uniqueness theorems}, Cambridge
University Press,
  Cambridge, 1996.

\bibitem{HorowitzMyers}
G.T. Horowitz and R.C. Myers, \emph{The {AdS/CFT} correspondence
and a new
  positive energy conjecture for general relativity}, Phys. Rev. \textbf{D59}
  (1999), 026005 (12 pp.).

\bibitem{Kasner}
E.~Kasner, \emph{Solutions of the {E}instein equations involving
functions of
  only one variable}, Trans.\ AMS \textbf{27} (1925), 155--162.

\bibitem{key485}
S.~Kobayashi, \emph{Transformation groups in differential
geometry}, Ergebnisse
  der Math., vol.~70, Springer-Verlag, New York, 1972.

\bibitem{Korotkin:1994dw}
D.~Korotkin and H.~Nicolai, \emph{A periodic analog of the
{S}chwarzschild
  solution},  (1994), gr-qc/9403029.

\bibitem{Exactsolutions}
D.~Kramer, H.~Stephani, M.~MacCallum, and E.~Herlt, \emph{Exact
solutions of
  {E}instein's field equations}, Cambridge University Press, Cambridge, 1980.

\bibitem{LafontaineRozoy2}
J.~Lafontaine and L.~Rozoy,  (2002), in preparation.

\bibitem{Lee:fredholm}
J.~Lee, \emph{Fredholm operators and {E}instein metrics on
conformally compact
  manifolds}, math.DG/0105046, 2001.

\bibitem{Linet}
B.~Linet, \emph{The static, cylindrically symmetric strings in
general
  relativity with cosmological constant}, Jour.\ Math.\ Phys.\ \textbf{27}
  (1986), 1817--1818.

\bibitem{Myers:1987rx}
Robert~C. Myers, \emph{Higher dimensional black holes in
compactified space-
  times}, Phys. Rev. \textbf{D35} (1987), 455--466.

\bibitem{Nomizu}
K.~Nomizu, \emph{On local and global existence of {K}illing vector
fields},
  Ann. Math. \textbf{72} (1960), 105--120.

\bibitem{Smale}
S.~Smale, \emph{{An infinite dimensional version of Sard's
theorem}}, Am.\
  Jour.\ Math. \textbf{87} (1965), 861--866.

\bibitem{Weinstein3}
G.~Weinstein, \emph{{$N$}-black hole stationary and axially
symmetric solutions
  of the {E}instein/{M}axwell equations}, Commun.\ Part.\ Diff.\ Eqs.
  \textbf{21} (1996), 1389--1430.

\bibitem{Winstanley:2001bs}
E.~Winstanley and O.~Sarbach, \emph{On the linear stability of
solitons and
  hairy black holes with a negative cosmological constant: The even-parity
  sector}, Class. Quant. Grav. \textbf{19} (2002), 689--724.

\end{thebibliography}
\end{document}